\documentclass[notitlepage,superscriptaddress,nofootinbib,tightenlines,preprintnumbers,twocolumn]{revtex4-2}
\usepackage{amsmath,verbatim,latexsym,amssymb,indentfirst,mathrsfs,mathtools,amsthm,bbm,bm,hyperref,url,cancel,subcaption,enumitem}
\usepackage[font=small,labelfont=bf,format=plain,justification=raggedright,singlelinecheck=false]{caption}
\usepackage{verbatim,indentfirst}
\usepackage[title,titletoc]{appendix}
\usepackage{hyperref}
\usepackage{CJKutf8}

\usepackage{graphicx}
\usepackage{epstopdf}

\renewcommand{\thesection}{\Roman{section}}
\renewcommand{\thesubsection}{\Roman{section} \Alph{subsection}}
\renewcommand{\thesubsubsection}{\Roman{section} \Alph{subsection} \arabic{subsubsection}}

\makeatletter
\def\p@subsection{}
\makeatother
\makeatletter
\def\p@subsubsection{}
\makeatother

\makeatletter
\newcommand\footnoteref[1]{\protected@xdef\@thefnmark{\ref{#1}}\@footnotemark}
\makeatother

\usepackage{soul}

\usepackage{color}
\usepackage[dvipsnames]{xcolor}
\usepackage[normalem]{ulem}

\newcommand{\ryd}{{\rm r}}

\newcommand{\hc}{ {\rm h.c.} }

\newcommand{\Tr}{{\rm Tr}}   

\newcommand{\LParen}{ \bm{(} }
\newcommand{\RParen}{ \bm{)} }

\newcommand*{\Set}[1]{\left\{  #1  \right\}}

\newcommand*{\bra}[1]{\langle #1\rvert}
\newcommand*{\ket}[1]{\lvert #1 \rangle}
\newcommand*{\braket}[2]{\langle #1 \lvert #2 \rangle}
\newcommand*{\ketbra}[2]{\lvert #1 \rangle\!\langle #2 \rvert}
\newcommand*{\expval}[1]{\left\langle  #1  \right\rangle}

\begin{document}

\begin{CJK}{UTF8}{gbsn}

\title{Proposals for experimentally realizing quantum-autonomous gates}

\author{José Antonio Marín Guzmán}
\email{marin@umd.edu}
\affiliation{Joint Center for Quantum Information and Computer Science, NIST and University of Maryland, College Park, MD 20742, USA}
\affiliation{Department of Physics, University of Maryland, College Park, MD 20742, USA}

\author{Yu-Xin Wang (王语馨)}
\affiliation{Joint Center for Quantum Information and Computer Science, NIST and University of Maryland, College Park, MD 20742, USA}

\author{Tom Manovitz}
\affiliation{Department of Physics, Harvard University, Cambridge, Massachusetts 02138, USA}

\author{Paul Erker}
\affiliation{Atominstitut, Technische Universität Wien, 1020 Vienna, Austria}
\affiliation{Institute for Quantum Optics and Quantum Information, IQOQI Vienna,
Austrian Academy of Sciences, Boltzmanngasse 3, 1090 Vienna, Austria}

\author{Norbert M. Linke}
\affiliation{Joint Quantum Institute, University of Maryland, College Park, MD 20742, USA}
\affiliation{Duke Quantum Center and Department of Physics, Duke University, Durham, NC 27701, USA}
\affiliation{National Quantum Laboratory (QLab), University of Maryland, College Park, MD 20742, USA}
\affiliation{Department of Physics, University of Maryland, College Park, MD 20742, USA}

\author{Simone Gasparinetti}
\affiliation{Department of Microtechnology and Nanoscience,
Chalmers University of Technology, SE-412 96 Göteborg, Sweden}

\author{Nicole~Yunger~Halpern}
\email{nicoleyh@umd.edu}
\affiliation{Joint Center for Quantum Information and Computer Science, NIST and University of Maryland, College Park, MD 20742, USA}
\affiliation{Institute for Physical Science and Technology, University of Maryland, College Park, MD 20742, USA}
\affiliation{Department of Physics, University of Maryland, College Park, MD 20742, USA}

\date{\today}

%
%
\begin{abstract}
Autonomous quantum machines (AQMs) execute tasks without requiring time-dependent external control. Motivations for AQMs include the restrictions imposed by classical control on quantum machines' coherence times and geometries. Most AQM work is theoretical and abstract; yet an experiment recently demonstrated AQMs' usefulness in qubit reset, crucial to quantum computing. To further reduce quantum computing's classical control, we propose realizations of quantum-autonomous gates on three platforms: Rydberg atoms, trapped ions, and superconducting qubits. First, we show that a Rydberg-blockade interaction or an ultrafast transition can quantum-autonomously effect entangling gates on Rydberg atoms. Passive lasers control these gates quantum-autonomously. One can perform \textit{Z} or entangling gates on trapped ions quantum-autonomously, by sculpting a linear Paul trap or leveraging a ring trap. Finally, circuit quantum electrodynamics can enable quantum-autonomous \textit{Z} and \textit{XY} gates on superconducting qubits. The gates can serve as building blocks for (fully or partially) quantum-autonomous circuits, which may reduce classical-control burdens.
\end{abstract}

{\let\newpage\relax\maketitle}
\end{CJK}

\section{Introduction}
\label{sec:Intro}

Autonomous quantum machines (AQMs) undertake tasks without relying on time-dependent external control~\cite{Mitchison19,MarinGuzman24}. Researchers have proposed AQMs including quantum refrigerators \cite{Linden10, Correa14}, engines \cite{Brunner12, GelbwaserKlimovsky14, Roulet17}, Maxwell demons \cite{Koski15, Strasberg18}, clocks \cite{Erker17, Schwarzhans21, Prech24}, and detectors \cite{Schwarzhans26},  as well as a molecular error-correcting machine \cite{Guerreiro21}. Natural AQMs include a molecular switch that enables vision~\cite{YungerHalpern20}. Artificial AQMs have recently been realized with trapped ions~\cite{Maslennikov19} and superconducting qubits~\cite{Aamir25, Sundelin24}. An autonomous quantum refrigerator reset a superconducting qubit to near its ground state~\cite{Aamir25}, as required for quantum computation~\cite{DiVincenzo00}.

Granting quantum machines (partial or full) autonomy is challenging but may improve performance and lower costs. Building nonautonomous quantum devices requires substantial funding, innovation, and patience~\cite{Preskill23}. Hence autonomy may seem like an unnecessary hurdle. Yet classical control limits quantum devices. For example, control wires limit the number of superconducting qubits that can fit on a chip. Classical wires also introduce heat and noise, suppressing coherence~\cite{Grumbling19, Krinner19}. Furthermore, lowering quantum devices' energy costs can enhance quantum advantages over classical counterparts~\cite{Auffeves22}. Macroscopic control equipment consumes orders of magnitude more energy than the quantum devices controlled. For instance, in superconducting experiments, room-temperature control electronics can consume nearly 3 kW ~\cite[Suppl.~Inf.]{Arute19}. This expense is comparable to the $\approx\,$10 kW required to power the dilution refrigerator  that cools the qubits~\cite[Suppl.~Inf.]{Arute19}. Meanwhile, the quantum processor dissipates an amount of power on the order of tens of  $\mu$W \cite{Jiang25}.\footnote{The superconducting processor sits in a mixing chamber whose cooling power approximately equals 20 $\mu$W. This number upper-bounds the processor's power dissipation.} In contrast, consider classically controlling devices. This task often costs an amount of work comparable, or negligible compared, to the devices' energy scales. For example, setting a thermostat takes little energy; the climate-control system requires much more. Freeing quantum devices from classical control may enhance efficiencies, as well as coherence times and scaling.

In defining AQMs, we follow~\cite{MarinGuzman24}. An AQM executes a task, potentially by harnessing multiple components. Quantum theory describes the machine usefully.\footnote{
Quantum theory describes nearly everything, but it provides unhelpful models of many objects. For example, classical mechanics models wrenches more usefully than quantum theory does.}
The microscopic Hamiltonian responsible for the AQM's mission remains constant, once one actuates the AQM (``presses go''). However, one can exert time-dependent external control in building and initializing the AQM. These rules echo a drone's autonomy: one constructs a drone and charges its battery using time-dependent control, perhaps by leveraging factory equipment. After activation, though, the drone delivers packages independently. Because an AQM's microscopic Hamiltonian remains constant, no external classical system spends thermodynamic work on accomplishing the machine's task~\cite{Vinjanampathy16, Woods23, Elouard23}. However, an external system may perform work on machine components that do not directly contribute to the machine's mission (\cite{MarinGuzman24} and Sec.~\ref{sec:traps}). AQMs differ from quantum machines controlled by autonomous classical systems such as classical artificial intelligence: no classical system controls any AQM (in any way that contributes directly to the AQM's mission). To distinguish the machine types, we sometimes use the term \emph{quantum-autonomous}.

Autonomous quantum computation has attracted interest recently. In one approach (dissipative quantum computation), a quantum system couples to a tailored environment~\cite{Verstraete09, 21_Lebreuilly_Autonomous, Ghasemian23, Zapusek23, Wang23}. The environment guides the system toward a steady state that encodes a computation's output. In an abstract-theory approach, researchers designed an autonomous quantum processing unit \cite{Meier26}. A quantum state encodes a program, which directs the processing unit to enact unitaries on the computational register. In another abstract-theory approach, Woods showed that quantum control can increase quantum gates' speeds (as functions of resources such as energy)~\cite{24_Woods_Quantum}. We complement these approaches, proposing experimental implementations of quantum-autonomous gates. 

We design quantum-autonomous gates for three platforms:  Rydberg atoms, trapped ions, and superconducting qubits. These protocols introduce no extra reliance on continuous-wave lasers. Section~\ref{sec:Rydberg} shows how to perform quantum-autonomous entangling gates on  {Rydberg} atoms, leveraging Rydberg-blockade interactions and ultrafast transitions. Section~\ref{sec:Ions} details implementations of $Z$ and entangling gates on trapped ions. These quantum-autonomous schemes rely on certain trapping potentials (sculpted linear Paul traps and ring Paul traps). Section~\ref{sec:Superconduct} describes quantum-autonomous $Z$ and $XY$ gates for superconducting qubits. These gates benefit from circuit quantum electrodynamics (QED). Section~\ref{sec:outlook} outlines directions for future work. 

Our gates can serve as building blocks for (fully or partially) quantum-autonomous circuits. Our focus on single-gate implementations mirrors that of, e.g.,  \cite{Molmer99, Sorensen99, Jaksch00, Duan04}. Concatenating such gates would yield quantum-autonomous circuits. Reference \cite{Meier26} describes an abstract concatenation strategy, whose implementation would vary across platforms and so requires further designing. Instead of implementing a fully autonomous quantum circuit, one could begin with a classically controlled quantum algorithm and replace some gates with quantum-autonomous ones. The partial quantum autonomy would alleviate some of the classical control burden.

The following notation appears across the paper. $\sigma_\alpha$ denotes the Pauli-$\alpha$ operator, wherein $\alpha \in \{x, y, z \}$. If $s$ denotes or indexes a system, $\sigma_\alpha^{(s)}$ represents that system's Pauli-$\alpha$ operator. The $\sigma_z$ eigenstate $\ket{0} \equiv \ket{\uparrow}$ corresponds to the eigenvalue 1; and $\ket{1} \equiv \ket{\downarrow}$, to $-1$. The raising and lowering operators have the forms
$\sigma_\pm \coloneqq \frac{1}{2} ( \sigma_x  \pm  i \sigma_y )$. We set $\hbar$ to 1.

 \section{Rydberg atoms}
\label{sec:Rydberg}

 {Rydberg}-atom quantum computation has advanced rapidly over the past few years~\cite{Saffman10,16_Saffman_Quantum,19_Adams_Rydberg}. Section~\ref{RAComp} reviews relevant features of the platform. Section~\ref{RAAut} describes quantum-autonomous gates based on Rydberg-blockade interactions and on ultrafast transitions. Appendix~\ref{Levine_Pichler} outlines a more challenging alternative, based on the  Levine–Pichler gate. It frees Rydberg-blockade gates from single-atom laser manipulations~\cite{Levine19}.  

\subsection{Background:  {Rydberg}-atom quantum computing}
\label{RAComp}

This subsection overviews relevant features of Rydberg-atom quantum computers. We describe the platform in Sec.~\ref{sec:raArch}, Rydberg-blockade controlled-$Z$ gates in  {Sec.~\ref{RydbergCPh},} and ultrafast controlled-$Z$ gates in Sec.~\ref{UltrafastNA}.

\subsubsection{Architecture of Rydberg-atom  {array}}
\label{sec:raArch}

A Rydberg atom is an atom in a high-energy electronic state.\footnote{
Colloquially, the term refers to an atom undergoing a protocol during which the atom (or similar atoms) is sometimes in a high-energy state. We adopt this usage.}
Such a \emph{Rydberg state} $\ket{\ryd}$ corresponds to a large principal quantum number, such as $n = 70$.
Commonly used elements include  {Rb, Cs, Yb, and Sr}~\cite{Bluvstein22, Saskin19, Shi22}. Two  {lower-energy} levels, labeled $\ket{0}$ and $\ket{1}$, form a qubit. 

 Rydberg atoms $A$ and $B$ can interact strongly under two sets of circumstances. First, suppose that an $A$ is in a Rydberg state. $A$ shifts the energy of $B$'s $\ket{1}$–$\ket{\ryd}$ transition. The shift exceeds the transition's Rabi frequency. If subject to a laser pulse resonant with the $\ket{1}$–$\ket{\ryd}$ transition, $B$ cannot jump to $\ket{\ryd}$. $A$ induces a \emph{Rydberg blockade}. Second, suppose that $A$ and $B$ are simultaneously excited to Rydberg states. The atoms couple strongly via a dipole–dipole \emph{Rydberg interaction}. 


%
\subsubsection{Rydberg-blockade controlled-$Z$ gate}
\label{RydbergCPh}

Using the Rydberg blockade, one can implement a controlled-$Z$ gate~\cite{Jaksch00,Saffman10}. Denote the control qubit by $C$ and the target qubit by $T$. The qubits are ideally in a joint state $\ket{\text{ct}}$ (wherein $\text{c,t} \in \{0,1,\ryd\}$) at each step of the protocol. Lasers can couple the state $\ket{1}_{C,T}$ to the Rydberg state $\ket{\ryd}_{C,T}$, as detailed below. Throughout this subsubsection, all pulses drive the $\ket{1}$–$\ket{\ryd}$ transition. The splitting between $\ket{0}$ and $\ket{1}$ is sufficiently large that pulses cannot affect any atom in $\ket{0}$. 

A three-pulse sequence effects a controlled-$Z$ gate:
\begin{enumerate}

    \item \label{item_Ryd_CZ_1}
    A $\pi$-pulse maps the control atom's $\ket{1}_{C}$ state to the Rydberg state: 
    $\ket{1}_{C}  \mapsto  \ket{\ryd}_{C}$.
    
    \item \label{item_Ryd_CZ_2}
    A $2\pi$-pulse maps the target atom's $\ket{1}_{T}$ state to the Rydberg state and back:
    $\ket{1}_{T}  \mapsto  \ket{\ryd}_{T}
    \mapsto  \ket{1}_{T}$. 
    
    \item  {Another $\pi$-pulse maps the control atom's Rydberg state back to $\ket{1}_{C}$: }
    $\ket{\ryd}_{C}  \mapsto  \ket{1}_{C}$. 
    
\end{enumerate}

The pulse sequence effects a controlled-$Z$ gate as follows. First, suppose the atoms begin in $\ket{00}$. All pulses are off-resonant with the $\ket{1}$–$\ket{\ryd}$ transition, so the atoms' state remains constant. Second, suppose the atoms begin in $\ket{10}$. Pulse 2 is off-resonant, so the target's state remains constant. Pulses 1 and 3 rotate the control's Bloch vector through a total angle of $2\pi$. The joint state acquires a $\pi$ phase shift: 
$\ket{10} \mapsto e^{i\pi}\ket{10}= -\ket{10}$. Third, the joint initial state $\ket{01}$ picks up the same shift: $\ket{01} \mapsto -\ket{01}$. Fourth, suppose the atoms begin in $\ket{11}$. Pulse 1 leaves the control qubit in $\ket{\ryd}$. The blockade prevents pulse 2 from exciting the target to $\ket{\ryd}$, so the target remains in $\ket{1}_T$. Pulse 3 returns the control to $\ket{1}_{C}$. The joint state acquires a $\pi$ phase shift: $\ket{11} \mapsto -\ket{11}$. Across all four initial states, the pulse sequence enacts a controlled-$Z$ gate, to within a global phase.

 {A Rydberg-blockade gate's time scale equals the $\ket{1}$–$\ket{\ryd}$ transition's inverse Rabi frequency. Such gates typically last for hundreds of nanoseconds \cite{Jaksch00}. A shorter, nanosecond-scale gate, described in Sec.~\ref{UltrafastNA}, does not require a Rydberg blockade~\cite{Chew22, Xu22, Mahesh25}. }

\subsubsection{Ultrafast entanglement generation}
\label{UltrafastNA}

Unlike the Rydberg-blockade gate, the ultrafast operation features two atoms simultaneously in Rydberg states~\cite{Chew22, Xu22, Mahesh25}.  {The atoms arrive in those states due to }
ultrashort laser pulses, which last for picoseconds or femtoseconds. The pulse durations are much shorter than the time scale of the energy shift caused by the Rydberg blockade. Thus, both atoms can reach their Rydberg states. The atoms then entangle via the Rydberg dipole–dipole interaction.

Chew \textit{et al.} implemented an ultrafast energy exchange~\cite{Chew22}.  {They excited two $^{87}$Rb atoms into
$\ket{\text{dd}} = \ket{43D, 43D}$, whereupon interactions caused oscillations with 
$\ket{\text{pf}} = \frac{1}{\sqrt{2}}(\ket{45P, 41F}+\ket{41F, 45P})$.
One oscillation maps $\ket{\text{dd}}$ to $-\ket{\text{dd}}$ in $\approx 2$ ns.\footnote{
The oscillation introduces the phase as follows. Denote by $J$ the strength of the coupling between $\ket{\text{dd}}$ and $\ket{\text{pf}}$. Over a time $t$, $\ket{\text{dd}}$ evolves to
$\cos(Jt)\ket{\text{dd}}-i\sin(Jt)\ket{\text{pf}}$. At $t = \pi/J \approx 2$ ns, the system returns to $\ket{\text{dd}}$, but having picked up a $\pi$ phase.}
This phase can enable a quantum gate: the computational basis could consist of two states in the $5S_{1/2}$ manifold \cite{Wilk10, Levine19}. A pulse sequence would map one computational-basis state to $\ket{\rm dd}$.
If the atoms began in a superposition of computational-basis states, the reported oscillation would introduce a relative phase. In the spirit of~\cite{Chew22}, we focus in Sec.~\ref{sec:Ultrafast} on implementing one portion of the gate protocol---the relative-phase portion---quantum-autonomously. }

\subsection{Quantum-autonomous entangling of  {Rydberg} atoms}
\label{RAAut}

Section~\ref{sec:AutRydBl} describes a quantum-autonomous implementation of the Rydberg-blockade controlled-$Z$ gate. In Sec.~\ref{sec:Ultrafast}, we propose quantum-autonomous ultrafast entanglement generation.

\subsubsection{Quantum-autonomous Rydberg-blockade gate}
\label{sec:AutRydBl}

Here, we show how to implement Rydberg-blockade controlled-$Z$ gates quantum-autonomously. First, we argue that passively mode-locked lasers (PMLLs) can operate quantum-autonomously. Second, we show that they can drive the Rydberg transition. Third, we show how they can generate a pulse sequence that effects a Rydberg-blockade gate. The final argument relies on PMLLs' ability to serve as autonomous quantum clocks prevalent in the AQM literature~\cite{Erker17}. 

The Rydberg-blockade gate requires three laser pulses (Sec.~\ref{RydbergCPh}). A laser pulses when a classical control system turns it on and off. This control precludes quantum autonomy. However, quantum autonomy allows for a fixed-frequency laser free from time-dependent control. PMLLs offer this possibility \cite{Silfvast04}. 

Mode-locked lasers combine longitudinal cavity modes of different frequencies. When the modes are phase-locked (when their phases coincide), they interfere constructively to produce high-intensity outputs. This technique is widely used to generate ultrashort pulses, which typically last for picoseconds or femtoseconds \cite{Innerhofer03}.

Many PMLLs require no external control to pulse. They contain saturable absorbers that let high-intensity waves through while absorbing low-intensity waves \cite{Silfvast04}. To elaborate on PMLLs' autonomy, we recall how lasing works \cite{Fox06}. A laser contains a gain medium, a large set of atoms that begin mostly excited (with population inversion). Extra photons enter the gain medium, stimulating emissions from a few atoms, which stimulate more emissions. When enough atoms de-excite, the lasing ends unless the gain medium undergoes repumping (to reinstate population inversion). Such repumping conventionally involves time-dependent external control incompatible with autonomy. Hence a PMLL can participate in an AQM if the gain medium's initial population inversion powers all the lasing.\footnote{Alternatively, the pumping mechanism might be another autonomous quantum system, such as a large negative-temperature quantum reservoir.}

Having argued that PMLLs can operate quantum-autonomously, we argue that they can enable Rydberg-atom gates: PMLLs can emit pulses of the appropriate duration, frequency, and regularity. First, we support the duration claim generally. Then, we support all three claims by analyzing an implementation of a Rydberg-blockade gate.

During a Rydberg-blockade gate, atoms transition between $\ket{1}$ and $\ket{\ryd}$ due to pulses that last for hundreds of nanoseconds \cite{Saffman10}. PMLLs can produce such pulses, despite their often producing ultrashort pulses \cite{Xia14, Liu15}. Also, a mode-locked laser emits a train of pulses quickly and stably \cite{Bartels08}. Therefore, one can form a long pulse by stringing together short pulses.
 
\begin{figure}[h]
    \centering
    \includegraphics[scale = 0.11]{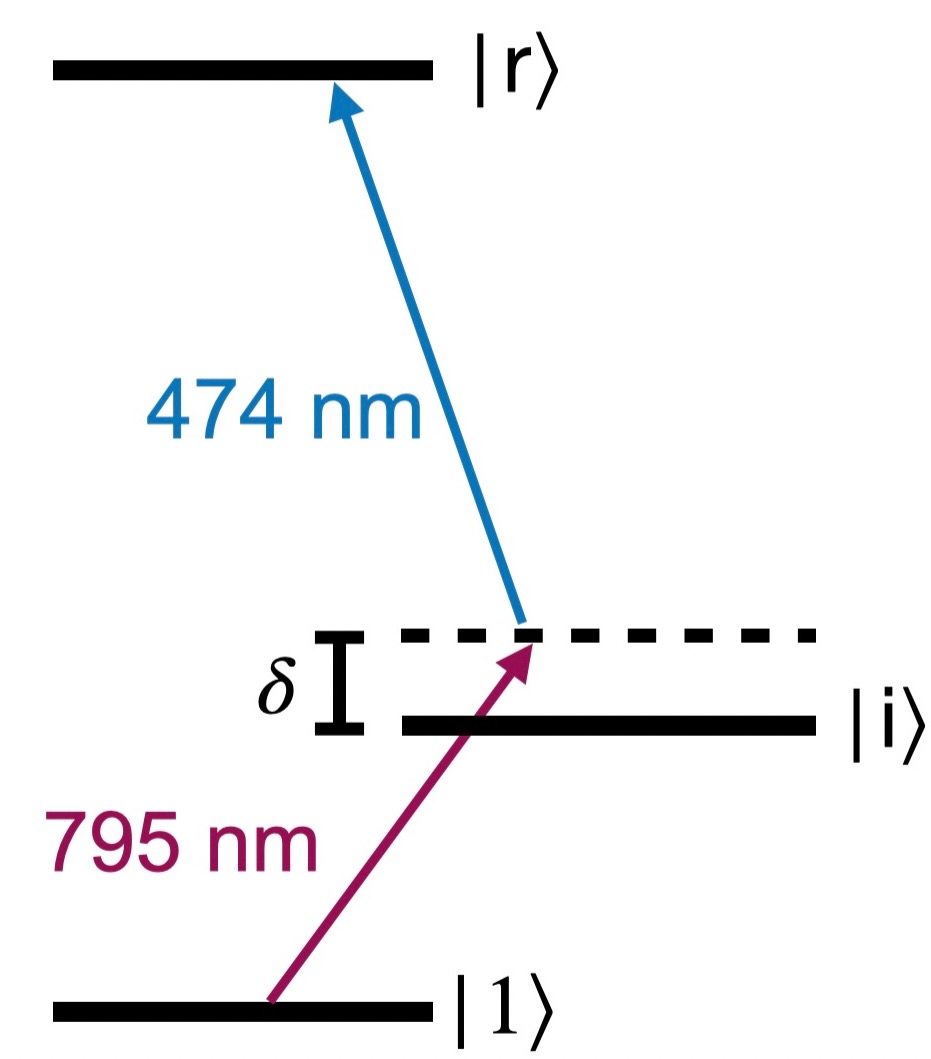}
    \caption{Energy-level diagram for a $^{87}$Rb atom (simplified from \cite{Gaetan09}). A two-photon process couples the qubit state $\ket{1}$ to the Rydberg state $\ket{\ryd}$. The lasers are detuned from $\ket{0}$–$\ket{1}$ and $\ket{1}$–$\ket{\rm r}$ transitions, so the process does not suffer from $\ket{\text{i}}$'s lossiness.}
    \label{fig:LevelDiagram}
\end{figure}

We now review an experimental implementation of a Rydberg-blockade gate and argue that PMLLs can provide pulses with the necessary characteristics (pulse duration, wavelength, and regularity) \cite{Gaetan09}. Figure~\ref{fig:LevelDiagram} illustrates a $^{87}$Rb atom's 
$\ket{1} = \ket{5S_{1/2}, F{=}2, m_F{=}2}$ and 
$\ket{\ryd} = \ket{58D_{3/2}, F {=} 3, m_F {=} 3}$ states in an energy-level diagram \cite{Gaetan09}. A two-photon process couples the states: $\ket{1}$ couples to the intermediate state $\ket{\text{i}} = \ket{5P_{1/2},F{=}2, m_F{=}2}$, which then couples to the Rydberg state.

Gaëtan \textit{et al}. performed a controlled-$Z$ gate on $^{87}$Rb atoms, using the pulse sequence described in Sec.~\ref{RydbergCPh} \cite{Gaetan09}. Two lasers implemented the $\ket{1}$–$\ket{\ryd}$ transition via a two-photon process (Fig.~\ref{fig:LevelDiagram} and step \ref{item_Ryd_CZ_1} in Sec.~\ref{RydbergCPh}). One laser was detuned from the $\ket{1}$–$\ket{\text{i}}$ gap, and the other laser from the $\ket{\text{i}}$–$\ket{\ryd}$ gap, by an amount $\delta$. ($\delta$ was large enough that the system never had a substantial population on $\ket{\text{i}}$, whose short lifetime would have threatened the atomic state's coherence.) The lasers' frequencies summed to the $\ket{1}$–$\ket{\ryd}$ gap. The lasers operated simultaneously, for approximately 70 ns \cite{Gaetan09}. PMLLs can emit pulses of such durations \cite{Xia14, Liu15}.

PMLL pulses can exhibit not only the durations, but also the wavelengths, needed to drive Rydberg transitions. A 795 nm pulse drives the $\ket{1}$–$\ket{\text{i}}$ transition \cite{Gaetan09}. Typical PMLLs have similar wavelengths, e.g., 780 nm to 2 $\mu$m~\cite{Bartels08, Zou20}. In contrast, a 474 nm pulse drives the $\ket{\text{i}}$–$\ket{\ryd}$ transition.  {To achieve such a short wavelength, one could employ a 948 nm PMLL and double the frequency using a nonlinear crystal \cite{Paschotta}. 
}

PMLLs can drive not only the $\ket{1}$–$\ket{\ryd}$ transition, but also a Rydberg-blockade gate. The nonautonomous gate protocol requires lasers to address different atoms at different times (Sec.~\ref{RydbergCPh}). Some clock must prompt each laser to turn on at the right instant. PMLLs can serve as autonomous quantum clocks, we argue next, aside from being able to drive the $\ket{1}$–$\ket{\ryd}$ transition.

A PMLL can act as a reliable \textit{autonomous quantum clock}, an AQM that regularly emits energy pulses called \emph{ticks} \cite{Erker17}.\footnote{
Autonomous quantum clocks are quantum-autonomous in the sense of Sec.~\ref{sec:Intro}. They are not atomic clocks~\cite{15_Ludlow_Optical} operated via autonomous classical control.}
PMLLs' laser pulses can serve as ticks. The pulses are regular \cite{Bartels08}, so PMLLs can keep time accurately. If driving a Rydberg transition, a PMLL keeps time for one atom. 

Another PMLL can keep time for a whole Rydberg-blockade protocol, we now show. For simplicity, we describe how a PMLL can time two other lasers that drive the Rydberg transition. The approach extends to more driving lasers, however. Suppose that transition-driving lasers $L_1$ and $L_2$ emit consecutive pulses. For example, $L_1$ can help perform the $\pi$-pulse in step~\ref{item_Ryd_CZ_1} of Sec.~\ref{RydbergCPh}; and $L_2$, the $2\pi$-pulse in step~\ref{item_Ryd_CZ_2}. $L_1$ should emit a pulse at time $t=0$; and $L_2$, at some time $t=T > 0$. An auxiliary laser $L_\text{a}$ can tell $L_2$ that an interval $T$ has elapsed since $L_1$ pulsed, according to the following protocol. Let $c$ denote the speed of light in a vacuum.\footnote{
Rydberg-atom platforms' environments are near-vacuums \cite{Bluvstein22, Ebadi21}.}
$L_\text{a}$ is set a distance $d = cT$ from $L_2$. One activates $L_1$ and $L_\text{a}$ simultaneously.  After a time $T$, $L_\text{a}$'s pulse reaches $L_2$. This pulse stimulates $L_2$ to fire.

We estimate the distance, and demonstrate its feasibility, as follows. Recall the controlled-$Z$ gate performed by Gaëtan \textit{et al}. \cite{Gaetan09}. A laser pulse drove the $\ket{1}$–$\ket{\ryd}$ transition over a duration $T \approx 70$ ns. Since $d = cT$, $L_\text{a}$ should lie a few meters from $L_2$. Several-meter-long optical fibers can delay light in experiments, so the control-clock scheme is feasible.

\subsubsection{Quantum-autonomous ultrafast entanglement generation}
\label{sec:Ultrafast}

Here, we propose quantum-autonomous version of ultrafast entanglement generation (Sec.~\ref{UltrafastNA}). We sketch the protocol first. We then argue that PMLLs can quantum-autonomously provide the necessary pulses, when augmented by passively Q-switched lasers or by cavities. 

The quantum-autonomous ultrafast gate, echoing the nonautonomous gate (Sec.~\ref{UltrafastNA}), proceeds as follows.  {First, a laser $L_1$ emits a picosecond-length pulse at both atoms. After roughly another picosecond, a laser $L_2$ emits another picosecond-length pulse, at an auxiliary laser's command. Together, $L_1$ and $L_2$ drive both atoms to Rydberg states, via transitions like the one depicted in Fig.~\ref{fig:LevelDiagram}.} The atoms interact via dipole-dipole coupling for a few nanoseconds. 
If $\approx 5~\mu$m separate the atoms, the state accumulates a conditional $\pi$ phase after $\approx 2$ ns~\cite{Chew22}. To halt the interaction, auxiliary lasers trigger pulses from additional lasers that de-excite both atoms.

Having outlined the protocol, we identify the laser resources needed to drive it, by specifying the atomic transitions involved. The experiment described in \cite{Chew22} involved three atomic levels of $^{87}$Rb: from least to greatest energy,
$\ket{\text{g}} = \ket{5S}$, $\ket{\text{e}} = \ket{5P}$, and the Rydberg state $\ket{\text{d}} = \ket{43D}$. 
A two-photon process (effected by two lasers) excites an atom from $\ket{\text{g}}$ through the intermediate $\ket{\text{e}}$ to $\ket{\text{d}}$. 
One laser is resonant with the $\ket{\text{g}}$–$\ket{\text{e}}$ transition. The atom must not remain in $\ket{\text{e}}$ for long, lest the level's population decay. Hence the second laser must turn on soon after the first (after a picosecond-scale interval). When in $\ket{\text{d}}$, two nearby atoms undergo the dipole–dipole interaction that enacts the gate.

Having specified the atomic transitions involved, we describe the laser resources required to drive them, demonstrating that PMLLs can fulfill the requirements above. A two-photon process drives the $\ket{\text{g}}$–$\ket{\text{d}}$ transition. The first step (the $\ket{\text{g}}$–$\ket{\text{e}}$ transition) requires a 780 nm pulse that lasts for  {$\approx 1$ ps.} ps. Pulse-to-pulse energy fluctuations limited the fidelity reported in \cite{Chew22}, on which \cite{Mahesh25} improved.
 
PMLLs, renowned for providing ultrashort pulses, can satisfy these requirements \cite{Bartels08}. The second step (the $\ket{\text{e}}$–$\ket{\text{d}}$ transition) requires a 480 nm pulse that lasts for 
 {$\approx 10$ ps.} A PMLL can attain this wavelength via frequency doubling (Sec.~\ref{sec:AutRydBl}). The pulse length falls within the range frequently achieved by PMLLs \cite{Bartels08}. 

However, time-dependent classical control augmented the lasers applied in the ultrafast protocol of \cite{Mahesh25}. That control, which included active mode locking, implemented the frequency conversion and amplification. Either of two strategies may eliminate the control. First, one could use a passively Q-switched laser. Such a laser contains an absorber that blocks low-power light from leaving the cavity. Once the power exceeds a threshold, the system outputs a short, energetic pulse \cite{Silfvast04, PaschottaQsw}. Passively Q-switched lasers can achieve the specifications necessary for driving ultrafast gates \cite{rpmc_nps_laser}:\footnote{Certain equipment, instruments, software, or materials are identified in this paper to specify the experimental procedure adequately. Such identification is not intended to imply recommendation or endorsement of any product or service by NIST; nor is it intended to imply that the materials or equipment identified are necessarily the best available for the purpose.} wavelengths of 480 nm and 780 nm (mentioned in the previous two paragraphs), a power of 5 W (used in \cite{Chew22}), and picosecond-length pulses.

Using the second strategy, one places the atoms in a cavity. One separates the mirrors such that light travels between them in the time between two laser pulses' arrivals. This matching amplifies the circulating pulses, which build up coherently. The primary challenge is the buildup time: the gradually strengthening pulses would drive the atoms undesirably. To circumvent this issue, one could keep the relevant atomic transition detuned from the laser. A pulse from another laser, separated by a delay line, would bring the atom into resonance after the buildup. These two strategies suggest the feasibility of autonomous control for ultrafast gates.

\section{Trapped ions}
\label{sec:Ions}

Here, we propose autonomous quantum gates for trapped ions. In Sec.~\ref{sec:common}, we overview conventional $Z$ and Mølmer–Sørensen (entangling) gates for trapped ions. We review ion traps, and argue that they do not preclude quantum autonomy, in Sec.~\ref{sec:traps}. Section~\ref{sec:IonGates} shows how tailored trapping potentials can enable quantum-autonomous $Z$ and Mølmer–Sørensen gates. Ring traps, too, can implement quantum-autonomous $Z$ gates, as we show in the same subsection.

\subsection{Background: Trapped-ion quantum computing}
\label{sec:common}
In trapped-ion quantum computing, an ion's 
internal (electronic) states serve as a qubit's logical states. Species used include $^{40}$Ca$^+$ \cite{Monz11}, $^{137}$Ba$^+$ \cite{Dietrich10}, $^{171}$Yb$^+$ \cite{Wright19}, $^{88}$Sr$^+$ \cite{Hughes20}, and $^9$Be$^+$ \cite{Sackett00}. For example, two states in a calcium ion's $S_{1/2}$ and $D_{1/2}$ manifolds can form a qubit.

In the most common geometry, the ions form a chain.\footnote{
One of our proposals involves a ring geometry, but discussing chains suffices here.}
It has normal vibrational modes that can transfer information between different ions' electronic degrees of freedom (DOFs). The modes correspond to a simple harmonic oscillator. Let $n$ denote the vibrational quantum number. We label the two ions' electronic states by $\uparrow$ and $\downarrow$.

\subsubsection{$Z$ gate}
\label{sec:SingQbTI}

In reviewing the trapped-ion $Z$ gate, we follow \cite{Schindler13}.\footnote{
An alternative scheme effects \emph{fast gates}, which require times much shorter than the ion motion's characteristic time scale \cite{Campbell10, Ospelkaus11}
Our quantum-autonomous gates rely on ion-trap features suited to longer interaction times. We therefore focus on standard single-qubit gates, which last for microseconds.} Consider an optical qubit encoded in a $^{40}$Ca$^+$ ion's $\ket{4S_{1/2}^{m_j =-1/2}}$ and $\ket{3D_{5/2}^{m_j =-1/2}}$ states.
Let $\Omega$ denote the ion transition's Rabi frequency. A $Z$ gate requires a laser whose frequency is detuned from $\Omega$ by some amount $\Delta \neq 0$. A laser wavelength of 729 nm serves. The laser shifts the ion's transition frequency by 
$\delta = \frac{\Omega^2}{2\Delta}$. Denote such a laser pulse's duration by $t$.
The shift leads to the unitary
$e^{-i\theta\sigma_z / 2} \, ,$ wherein $\theta = \delta t$. By driving the transition for 5.5 $\mu$s, experimentalists rotated a
qubit’s Bloch vector through an angle $\pi/2$ about the $z$-axis \cite{Schindler13}.

\subsubsection{Mølmer–Sørensen gate}

This section reviews the Mølmer–Sørensen gate, which entangles two ions by coupling their internal states to a shared vibrational mode \cite{Sorensen99, Molmer99}. Below, we overview the protocol and specify the laser resources required. We follow \cite{Molmer99}. 

The Mølmer–Sørensen gate unfolds as follows. Let $\omega_0$ denote each qubit's energy gap, $\nu$ a vibrational mode's frequency, and $\Delta$ the detuning from the motional sidebands. One illuminates the ions with two laser beams simultaneously. The beams have frequencies $\omega_0\mp \nu \pm \Delta$. Equivalently, one beam is detuned by $-\Delta$ from the red motional sideband; and the other, by $+\Delta$ from the blue: the beams are detuned around the carrier transition symmetrically.

Together, the two beams drive the two-photon transition $\ket{\downarrow\downarrow} \leftrightarrow \ket{\uparrow\uparrow}$, while transiently exciting the ions' motion. At a time $1/\Delta$, the motion is decoupled from the ions' spins. If the Rabi frequency has been chosen appropriately, the spins end up maximally entangled. 

In an experimental realization, the frequencies assumed the values $\omega_0 = 2\pi\times411$ THz, $\nu = 2\pi\times1.23$ MHz, and $\Delta = 20$ kHz~\cite{Benhelm08}. Operating for a time $1/\Delta \approx50\,\mu$s, the lasers prepared the entangled state $(\ket{\downarrow\downarrow}+i\ket{\uparrow\uparrow})/\sqrt{2}$.



\subsection{Ion traps and quantum autonomy}
\label{sec:traps}

In this subsection, we describe ion traps and their consistency with quantum autonomy. We focus on Paul traps, due to their commonness in quantum computing and for simplicity. 

Paul traps are the most popular ion traps applied in quantum computing. They confine ions using oscillating radio-frequency, as well as static, electric fields \cite{Bruzewicz19}. Paul traps shape ions into crystals whose geometries depend on the electrode arrangement. Four cylindrical electrodes can form a linear Paul trap. Alternatively, the electrodes can lie in a plane, trapping ions above it \cite{Kim10}. By using more planar electrodes, one can form a microfabricated surface trap that generates an arbitrary potential. Experimentalists have realized ring-shaped ion arrangements using such electrodes \cite{Li17}.

A Paul trap is a classical system whose electric fields oscillate to confine the ions. The oscillation drives a current, consuming and performing work. One might worry that this work prevents the platform from acting as an AQM: no classical control system may directly facilitate an AQM's mission by performing work, which would change the microscopic AQM Hamiltonian responsible for undertaking that mission~\cite{Vinjanampathy16,MarinGuzman24}. Yet the trap does not perform such work. Rather, the trap confines the ions by inducing a time-dependent potential. The potential does not operate on the information-bearing DOFs directly responsible for the machine's goal. Therefore, the quantum machine is autonomous.

Two objections suggest themselves. First, the trap consumes work to maintain the oscillating potential. Yet, as argued above, this work only confines the ions' positional DOFs radially. Similarly limited control facilitates our  Rydberg-atom and superconducting-qubit protocols, as detailed in App.~\ref{app_Control}. Second, as the ions oscillate, they push and pull the electrodes' electrons. The trap must counteract this effect to keep the voltage constant. Hence the trap draws work to increase or reduce the current. Yet this increase or reduction does not alter the ions' Hamiltonian. 
%

In summary, the trap does not prevent the ion system from acting as an AQM. Furthermore, quantum-autonomous lasers drive the gates, we argue in Sec.~\ref{sec:slide}. Hence the following gate protocols are quantum-autonomous, according to the definition in~\cite{MarinGuzman24}.

\subsection{Quantum-autonomous trapped-ion gates}
\label{sec:IonGates}

Section~\ref{sec:slide} shows how to implement $Z$ and Mølmer–Sørensen gates quantum-autonomously, by tailoring a potential to form a slide. Ring traps can enable quantum-autonomous $Z$ gates, too, we show in Sec.~\ref{sec:ring}.

\subsubsection{Tailored potential slide}
\label{sec:slide}

Our first trapped-ion proposal relies on a tailored potential. We describe the setup, its feasibility, and the extent of its quantum autonomy below. Then, we introduce the flash-lamp-pumped lasers that effect the gates. Finally, we argue for $Z$ and Mølmer–Sørensen gates' feasibility.

\begin{figure}[h]
    \centering
    \includegraphics[scale = 0.106]{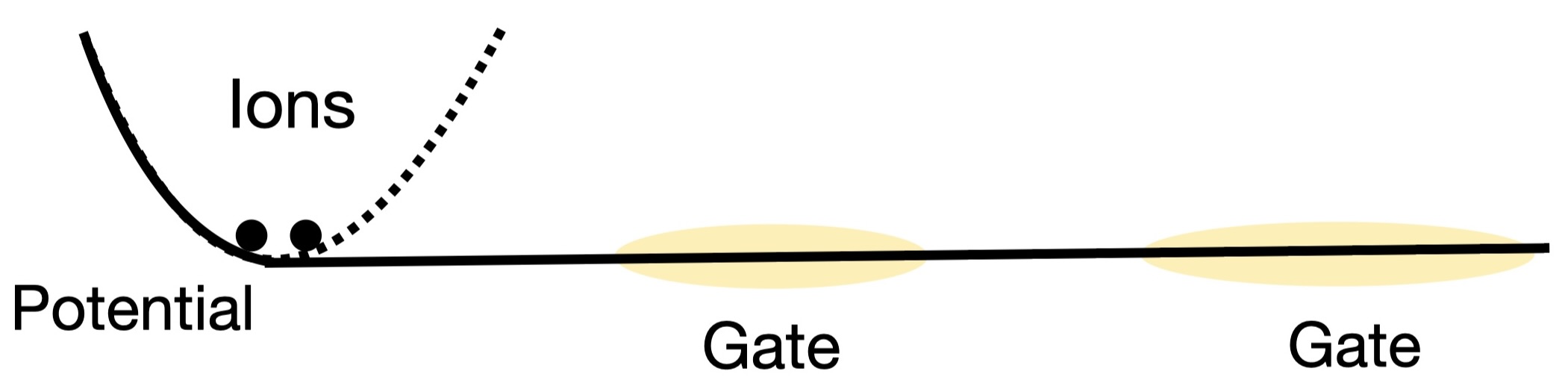}
    \caption{Tailored potential slide. A potential (dashed curve) confines the ions before the protocol, then is switched off. The ions initially slide down a potential gradient, then traverse the gate zones. Laser beams (yellow) implement gates.}
    \label{fig:Slide}
\end{figure}

The proposal features the following setup. Part of the trap forms a spatial gradient along which ions slide (Fig. \ref{fig:Slide}). Later, the trap levels out. Laser beams shine on particular locations. If an ion passes through a beam alone, it undergoes a single-qubit gate. If two ions pass through a beam simultaneously, they undergo a two-qubit gate.

The proposal involves a spatially varying potential and so an electric-field gradient. One can effect such a potential (or any other potential) using a Paul trap with an intricate-enough electrode structure \cite{Hong16}. Surface traps with multiple electrodes can serve this purpose  \cite{Seidelin06, Suleimen24}.  
Experimentalists have designed traps that feature many electrodes \cite{Sterk24}, rendering our proposal's requirements plausible. The potential ramp transports the ions axially. This transport does not affect the ions' internal states: the internal states couple, rather, to radial motional modes via lasers during entangling gates or only to the laser light during Z gates.\footnote{Gates have been realized via the shuttling of ions \cite{Leibfreid07, deClercq16, Tinkey22}. Active electrode control enabled this transport, in contrast with our quantum-autonomous proposal.} 


The gate is quantum-autonomous, consistently with Sec.~\ref{sec:traps}. The setup (laser placement and pumping, electrode arrangement, and loading of ions into the trap) occurs before the gate protocol. The ions are confined initially. When the protocol starts, the potential is switched off. This step amounts to ``pressing go" on the AQM. Without additional control, the ions roll down the potential slide (left-hand side of Fig. \ref{fig:Slide}). The ions then move along the flat potential, their center of mass maintaining a constant velocity. Coulomb repulsion increases the interatomic distance. The laser beams, detailed next, are applied throughout the protocol.

One final component of the setup needs detailing: the
flash-lamp-pumped lasers that effect the gates. A flash lamp is a gas-filled tube that emits light when triggered by a pulse \cite{Silfvast04}.\footnote{
The flash lamp is not the laser; rather, it pumps the laser: in flash-lamp pumping, one charges a capacitor. A trigger electrode discharges the capacitor, which energizes the lamp's gas. The excited gas flashes, pumping the laser's gain medium.
}
Flash-lamp-pumped lasers emit pulses that typically last for several microseconds to a few tens of microseconds \cite{PaschottaPG}. Hence these lasers can provide the 5.5–50~$\mu$s pulses required for trapped-ion gates.\footnote{
In contrast, PMLLs (the quantum-autonomous lasers discussed in Sec.~\ref{sec:AutRydBl}) emit picosecond- and nanosecond-scale pulses.
}
Industrial
flash-lamp-pumped lasers support wavelengths from 266
to 2,940 nm \cite{PaschottaLamp}. Such lasers can achieve the approximately 729 nm wavelengths required for the $Z$ and Mølmer–Sørensen gates \cite{Schindler13, Benhelm08}. Furthermore, one can charge a flash lamp's capacitor before implementing a gate. Therefore, flash-lamp-pumped lasers can enable quantum-autonomous control.  However, flash-lamp-pumped lasers suffer from poor frequency and power stability \cite{PaschottaLamp}. These shortcomings prevent the lasers, in their present form, from implementing high-fidelity gates. However, flash-lamp-pumped lasers have not been optimized for this purpose. Adapting them may enable high-quality quantum-autonomous control.

Having described the general setup, we detail quantum-autonomous $Z$ gates and Mølmer–Sørensen gates. First, we describe the $Z$ gate. Recall, from Sec. \ref{sec:SingQbTI}, that a  5.5 $\mu$s laser pulse implements a $Z$ gate. Our protocol can implement such a gate if an ion can remain in the laser beam for the required duration. Meeting this condition requires a correspondingly engineered trap. Although such a trap has not been realized, the intricate electrode structures required have been \cite{Suleimen24, Sterk24}, rendering this task feasible.

The slide can effect also a Mølmer–Sørensen gate. 
During it, two ions slide together. They pass through a bichromatic laser beam detuned from their qubit and radial (shared-motional-mode) frequencies. The shared mode transfers excitations between the ions' electronic DOFs.\footnote{If the trap design allows ions to move at a speed of $\approx 0.5$ m/s, a beam of diameter of $\approx 25$ $\mu$m can implement such an entangling gate. Appropriately distributed electrodes can give rise to this speed \cite{Suleimen24,Sterk24}. Standard linear-optics techniques can tune the beams' diameters.}

\subsubsection{Ring-trap autonomous gate}
\label{sec:ring}

Like a potential slide, a ring trap (Sec. \ref{sec:traps} and~\cite{Wang15, Li17, Urban19}) enables a quantum-autonomous $Z$ gate. First, we sketch the gate protocol. Then, we review specifications required in similar nonautonomous setups. These specifications, we argue, are feasible in our setup. 

\begin{figure}[h]
    \centering
    \includegraphics[scale = 0.10]{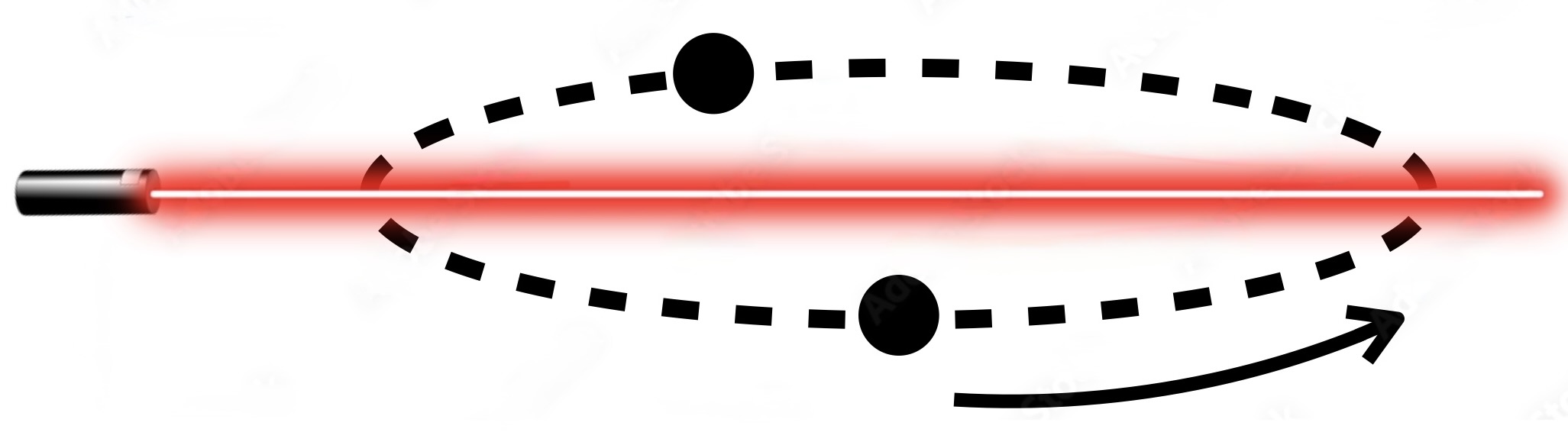}
    \caption{Quantum-autonomous $Z$ gate implemented with a ring trap. The black disks represent ions. Their trajectory forms the dashed line. A laser shines on two ring segments, driving internal transitions in the ions.}
    \label{fig:Ring}
\end{figure}

Figure \ref{fig:Ring} illustrates the quantum-autonomous gate. The ions traverse a ring.\footnote{We picture multiple ions, despite describing a single-qubit gate for the convenience of our feasibility analysis.} A  laser beam illuminates two segments of the ring.
The ions pass into and out of the illuminated  segments several times, until the laser is discharged, as in Sec.~\ref{sec:slide}.

The ions can revolve quickly enough to effectively experience one coherent pulse throughout the protocol. Consider a $^{43}\text{Ca}^+$ ion with qubit states 
$\ket{\downarrow} =\ket{4S_{1/2}^{F{=}4,M_F{=}0}}$ and 
$\ket{\uparrow} =\ket{4S_{1/2}^{F{=}3,M_F{=}0}}$, following  \cite{Ballance16}. The desired pulse induces a relative phase between the ion state's $\ket{\downarrow}$ and $\ket{\uparrow}$ terms. Passing through an illuminated arc, the ion experiences a short pulse. Suppose the qubit decoheres much more slowly than the ion revolves. The short pulses jointly resemble one long pulse, increasingly shifting the qubit's relative phase. 

Now, we support the supposition that the qubit decoheres much more slowly than the ion revolves. Reference \cite{Urban19} reports on an experiment in which two calcium ions revolved at a frequency of 100 kHz. In contrast, a calcium qubit decoheres at a rate $0.167$ Hz \cite{Ballance16}. As 0.167 Hz $\ll$ 100 kHz, the time scales meet the required condition.

\section{Superconducting qubits}
\label{sec:Superconduct}

This section introduces quantum-autonomous gates for transmon qubits in circuit QED. Using circuit-QED, one typically controls gates using microwave tones. Our proposals, in contrast, rely on quantum-autonomous photon sources, described below. Sections~\ref{sec:SCz} and~\ref{sec:PhaseShift} describe quantum-autonomous $Z$ gates. In the protocol of Sec.~\ref{sec:SCz}, a transmon-and-cavity system reflects an incoming photon. The photon is resonant with the transmon's ground-to-first-excited-state transition. Therefore, the photon alters the relative phase between the transmon's two lowest energy levels. In the protocol of Sec.~\ref{sec:PhaseShift}, an incoming photon enters a cavity coupled to a transmon dispersively. Incrementing the cavity's population, the photon effectively changes the qubit's gap (via the dispersive coupling) and so changes the transmon's relative phase. Section~\ref{sec:XY} describes a quantum-autonomous $XY$ (entangling) gate.

\subsection{$Z$ gate effected via reflected photon in circuit QED}
\label{sec:SCz}

In Sec.~\ref{sec:OrdCZ}, we review a nonautonomous protocol implemented by Besse \emph{et al.}~\cite{Besse18}. During the protocol, a transmon–cavity system reflects an incoming photon. The protocol effects a controlled-$Z$ gate on the photon and transmon. On page 2 of \cite{Besse18}, Besse \emph{et al.} cast the gate as controlled on the transmon and targeting the photon. Yet one can interpret the gate as controlled on the photon and targeting the transmon. Privileging the latter interpretation, suppose the photon is in $\ket{1}$ (rather than in a nontrivial superposition of $\ket{0}$ and $\ket{1}$). The gate enacts a $Z$ on the transmon. We show how to implement this gate autonomously in Sec.~\ref{sec:autcz}.

\subsubsection{Nonautonomous gate effected via reflected photon in circuit QED}
\label{sec:OrdCZ}

\begin{figure}[h]
    \centering
   {\includegraphics[width=\linewidth]{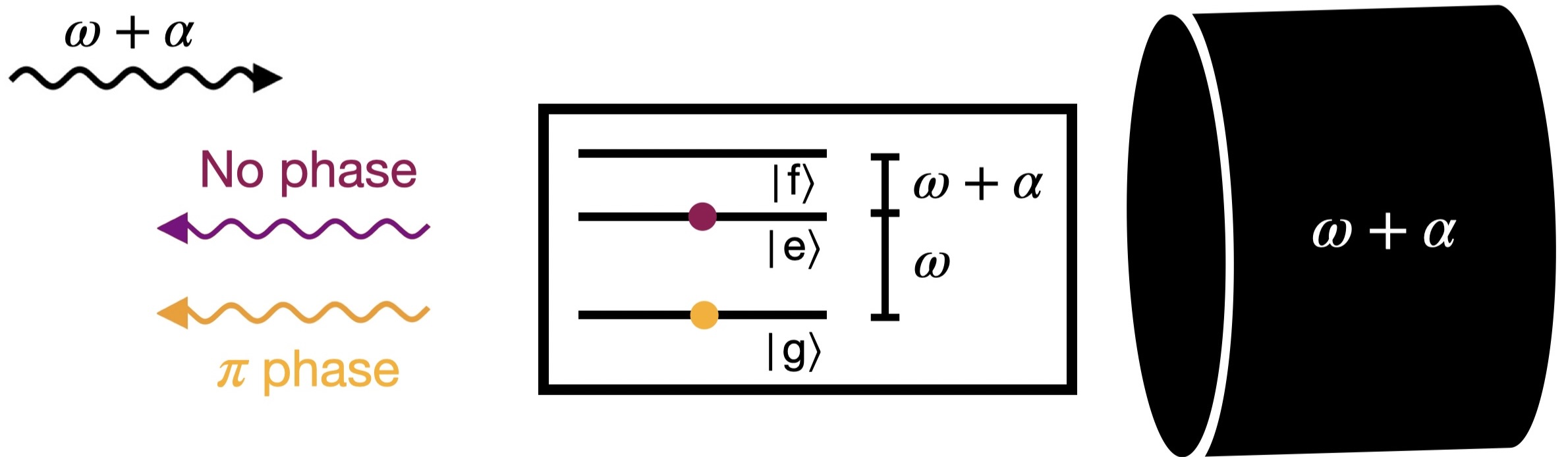}}
    \caption{Controlled-$(-Z)$ gate implemented in circuit QED. An incoming photon interacts with a transmon–cavity system, which reflects the photon. The photon acquires a phase dependent on the transmon's initial state; equivalently, the transmon acquires a phase dependent on the photon's initial state. Figure simplified from~\cite{Besse18}. 
    }
    \label{fig:CZgate}
\end{figure}

Here, we review the controlled gate reported on in~\cite{Besse18}. The protocol features a transmon coupled to a single-mode cavity (Fig.~\ref{fig:CZgate}). The transmon has a ground state $\ket{\text{g}} \equiv \ket{0}$, first excited state $\ket{\text{e}} \equiv \ket{1}$, and second excited state $\ket{\text{f}}$. Let $\omega$ denote the $\ket{\text{g}}$–$\ket{\text{e}}$ energy gap. The transmon has an anharmonicity $\alpha$; a gap $\omega+\alpha$ separates the $\ket{\text{e}}$ and $\ket{\text{f}}$ energies. The cavity's electromagnetic mode, too, has a frequency $\omega+\alpha$. We denote the Fock states by $\ket{n}$, wherein $n = 0, 1, \ldots$ 

The gate protocol proceeds as follows. One shoots a photon of frequency $\omega + \alpha$ at the transmon–cavity system. The subsequent events depend on the transmon’s state.

First, suppose the transmon begins in $\ket{\text{g}}$. Due to its energy spectrum, the transmon cannot interact with the photon. The cavity interacts instead: it reflects the photon, imparting a relative phase of $\pi$. If the photon began in 
$a \ket{0} + b \ket{1}$ (wherein $a, b \in \mathbb{C}$), it ends in
$a \ket{0} - b \ket{1}$.

Now, suppose the transmon begins in $\ket{\text{e}}$. The photon is resonant with the transmon's $\ket{\text{e}}$–$\ket{\text{f}}$ transition. This transition, however, couples to the cavity mode via a Jaynes–Cummings interaction. The coupling strength exceeds the cavity linewidth. Hence the incoming photon has the wrong frequency to interact with the transmon–cavity system. The system reflects the photon without imparting any phase to it. 

Let us summarize the gate's action. Suppose the photon begins in 
$a \ket{0} + b \ket{1}$. We have seen that
\begin{align}
   & \ket{\rm g} ( a \ket{0} + b \ket{1} )
   \mapsto  \ket{\rm g}  ( a \ket{0} - b \ket{1} ) 
   \quad \text{and} \\
   & \ket{\rm e} (a \ket{0} + b \ket{1} )
   \mapsto  \ket{\rm e} ( a \ket{0} + b \ket{1} ) .
\end{align}
Suppose the transmon begins in $\alpha \ket{\rm g} + \beta \ket{\rm e}$, wherein $\alpha, \beta \in \mathbb{C}$. The protocol acts as
\begin{align}
   & (\alpha \ket{\rm g} + \beta \ket{\rm e} )
   (a \ket{0} + b \ket{1} ) \\
   & \mapsto a ( \alpha \ket{\rm g} + \beta \ket{\rm e} ) \ket{0}
   + b ( - \alpha \ket{\rm g} + \beta \ket{\rm e} )  \ket{1} . 
\end{align}
The photon has controlled a $-Z$ gate on the transmon. Now, suppose the photon begins in $\ket{1}$, as throughout the following subsubsection. The protocol transforms the transmon state as
$\alpha \ket{\rm g} + \beta \ket{\rm e}
\mapsto  - \alpha \ket{\rm g} + \beta \ket{\rm e}
=  -Z ( \alpha \ket{\rm g} + \beta \ket{\rm e} )$.
The global phase $-1$ lacks physical significance, so the protocol enacts a $Z$ gate.

\subsubsection{Quantum-autonomous $Z$ gate effected via reflected photon in circuit QED}
\label{sec:autcz}

Here, we show how to quantum-autonomously implement the $Z$ gate described in the previous subsubsection. We describe the setup, then detail the protocol. The primary challenge is to stop the gate from acting repeatedly; our solution relies on a $\Lambda$ system.

The protocol features the following setup. A transmon $A$ couples to a cavity via the Jaynes–Cummings interaction described in Sec.~\ref{sec:OrdCZ}. The cavity's electromagnetic mode is resonant with an autonomous quantum clock's ticks. The clock couples to the cavity via a waveguide. The clock consists not of a PMLL, as in Sec.~\ref{sec:AutRydBl}, but of several qubits and a multilevel qudit~\cite{Schwarzhans21,Meier25}. The qubits form a clockwork mechanism that contacts hot and cold baths. From the temperature gradient, the clockwork extracts thermodynamic work. This work drives the qudit to its topmost energy level. The qudit then emits a photon, interpretable as a tick, while dropping to its ground state. The clock can achieve a high accuracy and precision if the clockwork is sufficiently complex~\cite{Schwarzhans21}. 

The gate proceeds as follows. The autonomous quantum clock emits a photon. The photon interacts with the transmon–cavity system, which reflects it, as described in Sec.~\ref{sec:OrdCZ}. The interaction implements a $Z$ gate on the transmon.

Having described the protocol, we address a challenge to it: by definition, a clock ticks more than once~\cite{Erker17,Schwarzhans21}. Whenever the gate-controlling clock ticks, the gate happens again. Our solution centers on a metastable state. Appendix~\ref{app_SC_frac} sketches an alternative, in which the clock applies a fraction of the gate many times.

\begin{figure}[h]
    \centering
   {\includegraphics[width=0.89\linewidth]{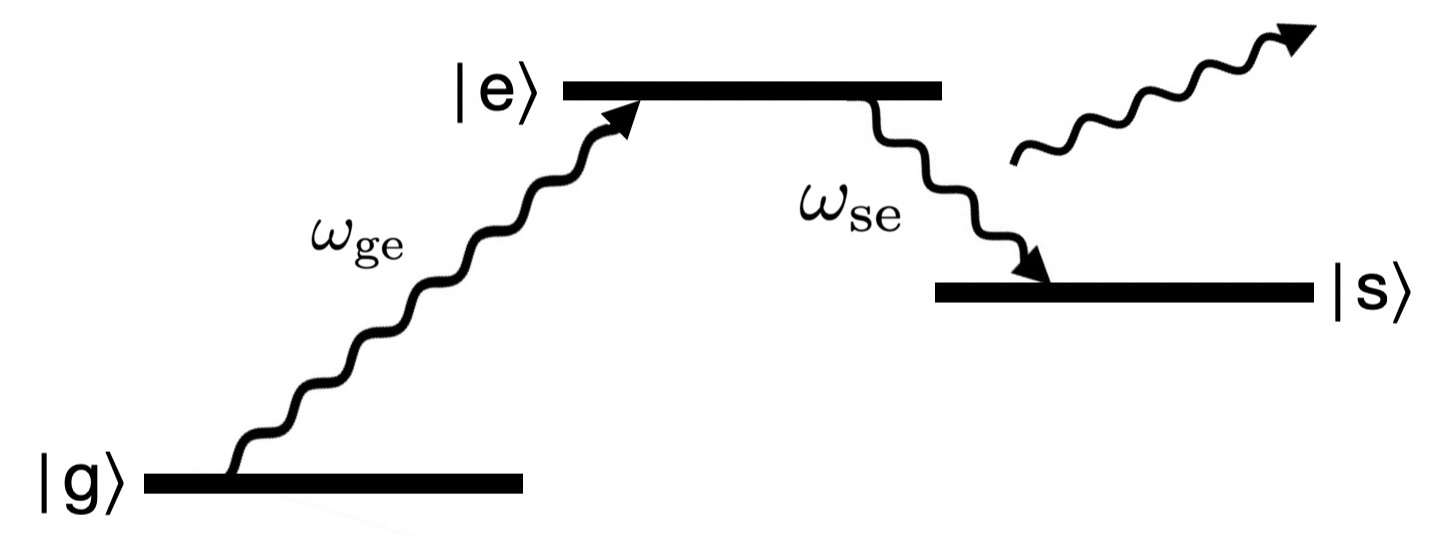}}
    \caption{$\Lambda$ system in an autonomous quantum clock. A clockwork mechanism (Fig.~\ref{fig:engine}) pumps the system into $\ket{\text{e}}$.  The system spontaneously emits frequency-$\omega_\text{se}$ photons, which serve as clock ticks.}
    \label{fig:sState}
\end{figure}

The clock can have a metastable state if it contains a three-level $\Lambda$ system~\cite{Lambropoulos07}, as has been engineered from a transmon \cite{Novikov16, Inomata16}. Figure~\ref{fig:sState} depicts the $\Lambda$ system. $\ket{\text{g}}$ denotes the ground state; and $\ket{\text{e}}$, an excited state. $\ket{\text{s}}$ denotes a metastable intermediate-energy level. $\omega_\text{ge}$ denotes the $\ket{\text{g}}$–$\ket{\text{e}}$ energy gap; and $\omega_\text{se}$, the $\ket{\text{e}}$–$\ket{\text{s}}$ gap.

The $\Lambda$ system replaces the multilevel qudit in an autonomous quantum clock (Sec.~\ref{sec:AutRydBl} and~\cite{Schwarzhans21,Meier25}). The $\Lambda$ system begins in $\ket{\text{g}}$. Absorbing an excitation from the clockwork (detailed in the next paragraph), the $\Lambda$ system transitions to $\ket{\text{e}}$. It then decays into $\ket{\text{s}}$, emitting a frequency-$\omega_\text{se}$ photon. This tick initiates a $Z$ gate. Suppose the $\ket{\text{s}}$ lifetime far exceeds the gate-operation time. The $\Lambda$ system likely does not decay from $\ket{\text{s}}$ to $\ket{\text{e}}$, and tick again, until after the gate ends. This separation of time scales is feasible: in~\cite{Besse18}, the optimal gate interaction lasted for $\approx 250$ ns. This duration is far shorter than the lifetime exhibited by a transmon $\Lambda$ system's metastable state, $\approx 4$ $\mu$s \cite{Novikov16}. 

\begin{figure}[h!]
    \centering
        \centering
        \includegraphics[scale=0.115]{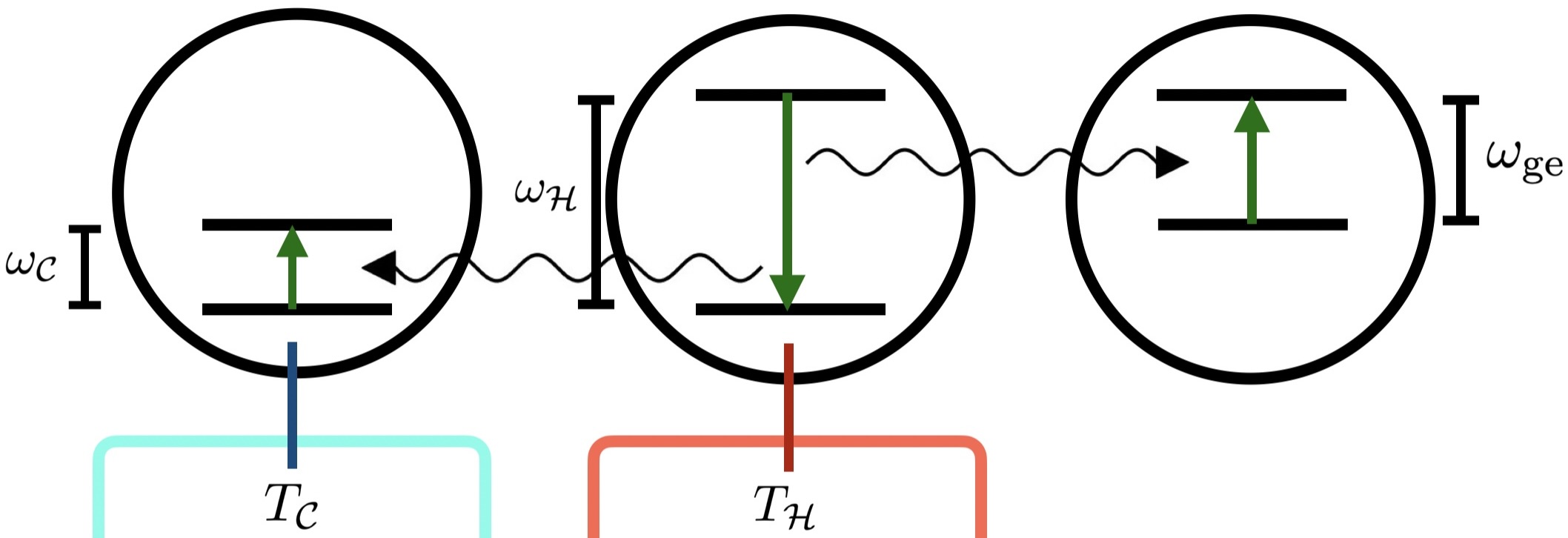}
        \caption{Autonomous quantum clock featuring a $\Lambda$ system driven by an autonomous quantum engine. The resonance condition $\omega_\mathcal{C} + \omega_\text{ge} = \omega_\mathcal{H}$ enables a three-body energy exchange. The thermal baths favor the $\ket{01\text{g}} \mapsto \ket{10\text{e}}$ transition, driving excitations into the $\Lambda$ system (Fig.~\ref{fig:sState}).
        }
        \label{fig:engine}
\end{figure}

An autonomous quantum heat engine can drive the $\Lambda$ system's $\ket{\text{g}}$–$\ket{\text{e}}$ transition, via a three-body interaction~\cite{Erker17, Schwarzhans21}. Figure~\ref{fig:engine} depicts this clock. A cold qubit $\mathcal{C}$ has a Hamiltonian
$H_\mathcal{C}=\frac{1}{2} \omega_\mathcal{C} \sigma_z^{(\mathcal{C})}$ characterized by a gap $\omega_\mathcal{C}$. This qubit, contacting a temperature-$T_\mathcal{C}$ bath, begins in the thermal state
$\rho_\mathcal{C} = e^{-H_\mathcal{C}/T_\mathcal{C}}/\Tr( e^{-H_\mathcal{C}/T_\mathcal{C}})$. The hot qubit $\mathcal{H}$ has analogous properties $H_\mathcal{H}$, $\omega_{\mathcal{H}}$, $\rho_\mathcal{H}$, and $T_\mathcal{H} > T_\mathcal{C}$.
Denote by $\ket{jk l}$ the state in which the $\Lambda$ system is in $\ket{j}$; $\mathcal{C}$, in $\ket{k}$; and $\mathcal{H}$, in $\ket{l}$.
Under the resonance condition $\omega_\mathcal{C} + \omega_\text{ge} = \omega_\mathcal{H} $, the qubits undergo the three-wave-mixing process
$\ket{01\text{g}}\leftrightarrow\ket{10\text{e}}.$
The heat baths favor the rightward direction, exciting the $\Lambda$ system and $\mathcal{C}$. $\mathcal{C}$ can couple strongly to its bath, which dissipates the excitation quickly~\cite{Aamir25}. This dissipation renders the clock's evolution irreversible.
One can introduce more $\mathcal{CH}$ pairs, and make the $\Lambda$ system more elaborate, to improve the clock's accuracy and precision~\cite{Schwarzhans21}.

\subsection{Quantum-autonomous $Z$ gate implemented via dispersive coupling in circuit QED}
\label{sec:PhaseShift}

We have introduced a quantum-autonomous $Z$ gate implemented via a reflected photon; now, we introduce a quantum-autonomous $Z$ gate implemented via a dispersive coupling. The qubit is a transmon interacting dispersively with a cavity. An autonomous quantum clock emits a photon that enters the cavity via a waveguide. As the cavity's photon number shifts, so does the transmon's effective frequency, due to the dispersive coupling. Hence the transmon's state acquires a relative phase. Below, we detail the protocol and argue for its feasibility.

Our protocol resembles the Duan–Kimble gate~\cite{Duan04}. There, a photon interacts with a transmon–cavity system to perform a logic operation on the transmon. This scheme has been experimentally implemented in cavity QED \cite{Hacker16} and circuit QED \cite{Besse18, Kono18}.

In our proposal, the transmon couples to the cavity dispersively with a strength $\chi$. Let $a$ denote the cavity mode's lowering operator; and $\sigma_z$, the transmon's Pauli-$z$ operator. The transmon–cavity Hamiltonian contains an interaction term 
$\frac{\chi}{2} a^\dagger a \, \sigma_z$.
Since the interaction is dispersive, the transmon's $\ket{0}$ and $\ket{1}$ populations remain constant. In contrast, the cavity population decays at a rate $\gamma$. 

We model the incoming photon as a Gaussian pulse of bandwidth $\Omega$. The bandwidth is narrow, compared to the cavity's decay rate, which is lower than the transmon–cavity's coupling strength: $\Omega\ll\gamma\ll\chi$. In this regime, the transmon state (expressed relative to the computational basis) acquires the relative phase~\cite{Besse18}
\begin{align}
    \phi = 2\arctan\left( 2\chi / \gamma \right).
    \label{eq:Shift}
\end{align}
Upon beginning in $a \ket{0} + b \ket{1}$ (wherein $a, b \in \mathbb{C}$), the transmon ends in $a \ket{0} + e^{i \phi} \, b \ket{1}$.

Appendices~\ref{app_EOMs} and \ref{app_Fidelity} report on numerical simulations of the gate. The transmon suffers from very little dephasing in the narrow-bandwidth, strong-coupling regime: in the simulation, the qubit's Bloch vector rotates through an angle $\pi/2$ with a fidelity of 0.9989. A coherence measure, defined in App.~\ref{app_Fidelity}, simultaneously achieves a value of 0.4997 (the maximum possible value is 0.5000). 

Furthermore, the gate lasts for a reasonable time. Appendix~\ref{app_Fidelity} shows numerically that the system reaches a steady state by a time $\approx 200/\gamma$. Typical cavity lifetimes $1/\gamma$ are on the order of 10 ns \cite{Krantz19}. Hence the gate operates for a time $\approx 2$ $\mu$s. In contrast, transmon qubits can have dephasing times $T_2^*$ on the order of 100 $\mu$s~\cite{Krantz19}. The gate can therefore end well before the qubit decoheres.

\subsection{Quantum-autonomous \textit{XY} gate in circuit QED}
\label{sec:XY}

This section shows how to implement a quantum-autonomous $XY$ gate on transmons. First, we review the $XY$ gate's definition and transmons' interactions. Then, we describe the setup and protocol. We argue for the protocol's feasibility next. Finally, we describe two challenges and resolutions to them.

We define the $XY$ gate as 
\begin{align}
    U_{XY}(\theta) \coloneqq e^{-i\frac{\theta}{4}\,(\sigma_x \otimes \sigma_x+\sigma_y \otimes \sigma_y)}.
    \label{eq:XYgate}
\end{align}
 $\theta$ denotes the rotation angle in the subspace $\text{span}\{\ket{01},\ket{10}\}$. 
 $U_{XY}(\theta)$ fully or partially transfers an excitation between two qubits (a transmon's native Hamiltonian being $\propto \sigma_z$). $XY$ interactions are native to solid-state-qubit platforms~\cite{Abrams20}. One can couple transmons using microwave drive tones, by parametrically modulating a coupler element (e.g., a flux-tunable superconducting quantum interference device), or by dynamically bringing the transmons into resonance using a flux pulse. Our protocol couples transmons by tuning one into resonance with another quantum-autonomously.

\begin{figure}[h]
    \centering
    \includegraphics[width=\linewidth]{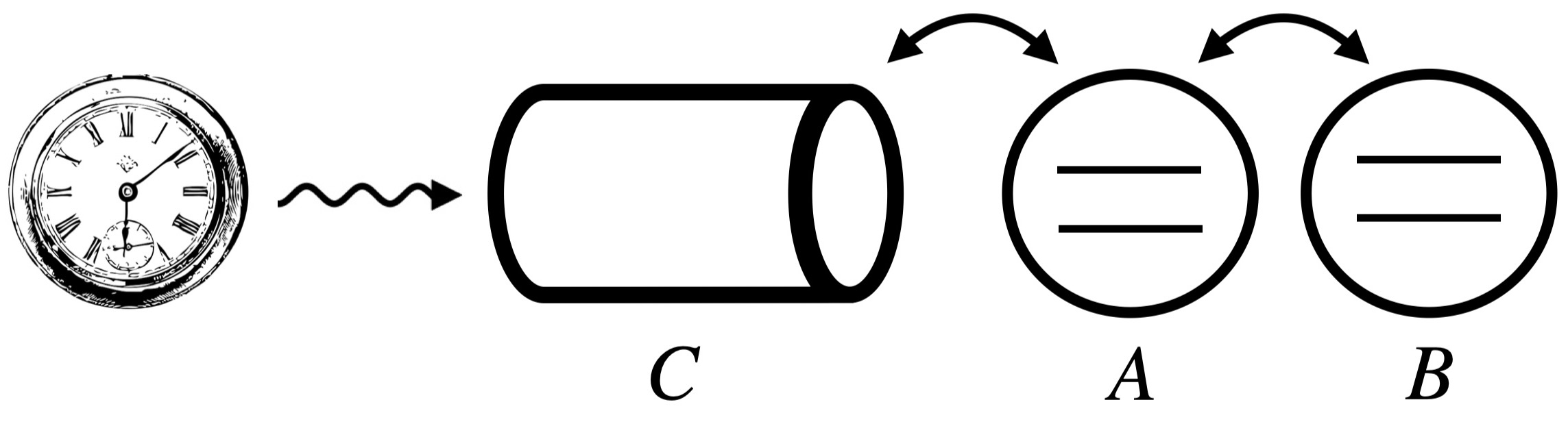}
    \caption{Quantum-autonomous implementation of $XY$ gate in circuit QED. An autonomous quantum clock emits a photon into a cavity $C$ coupled dispersively to transmon $A$. Via that coupling, the photon effectively changes $A$'s frequency. If $A$ becomes resonant with $B$, the transmons interact.
    }
    \label{fig:QbCav}
\end{figure}

Figure~\ref{fig:QbCav} depicts the setup. A transmon $A$ has a tunable bare frequency $\omega_A$. $A$ couples capacitively, with a strength $g_{AC}$, to a cavity $C$. The cavity supports an electromagnetic mode of frequency $\omega_C$. We aim to couple $A$ to a transmon $B$, which has a fixed frequency $\omega_B$. 

The gate proceeds as follows. As in Sec. \ref{sec:SCz}, an autonomous quantum clock ticks. The tick photon enters the cavity. Via the cavity–transmon coupling, the photon effectively shifts transmon $A$'s frequency. If $A$'s effective frequency matches $B$'s frequency, the transmons interact.

 The tick photon changes $A$'s frequency as follows. Let 
$\Delta_{AC}  \coloneqq \omega_A - \omega_C$ denote the difference between $A$'s and the cavity's frequencies. $A$ couples to the cavity dispersively. Denote the cavity's annihilation operator by $a$. The $CAB$ Hamiltonian follows from second-order perturbation theory \cite{Krantz19}:
\begin{align}
   \label{eq:DispH}
   H_{CAB }
   & = \omega_C  \,  a^\dagger a
   + \frac{1}{2}  \,  \omega_A\sigma_z^{(A)}
   + \chi_{AC}  \,  a^\dagger a  \,  \sigma_z^{(A)} 
   \\ \nonumber & \quad \,
   +\frac{1}{2}  \,  \omega_B  \,  \sigma_z^{(B)}
   +g_{AB}  \left( \sigma_+^{(A)}  \sigma_-^{(B)}  +  \hc  \right).
\end{align}

Due to the third term, the cavity's occupation number influences $A$'s effective frequency. Define $\Delta_{AB} \coloneqq \omega_B-\omega_A$ as the difference between the transmons' frequencies. Suppose that $\chi_{AC} = \Delta_{AB}$ and that the cavity begins empty. The first time the clock ticks, the cavity acquires one photon, which tunes $A$ into resonance with $B$. 

Now, we argue that $\chi_{AC}$ can equal $\Delta_{AB}$ in practice. This condition implies that the transmons' bare frequencies are close together: dispersive-shift constants $\chi_{AC}$ are usually much smaller than bare qubit frequencies $\omega_{A,B}$; and $\chi_{AC}$ must translate $\omega_A$ onto $\omega_B$. This closeness presents a challenge: Eq.~\eqref{eq:DispH} describes the interactions relevant in the presence of cavity photons. However, the cavity begins empty, and $A$ begins off-resonant with $B$. Under these circumstances, $A$ couples to $B$ via a $ZZ$ interaction. To quantify this interaction, we denote by $g_{{AB}}$ the strength of the $XY$ coupling between $A$ and $B$. When the cavity lacks photons, the qubits interact via a Hamiltonian term
$\frac{g_{AB}^2}{\Delta_{AB}} \sigma_{z}^{(A)} \sigma_{z}^{(B)}$ \cite{Mundada19}. The $XY$ coupling must dominate over this $ZZ$ coupling: 
$g_{AB} \gg g_{AB}^2 / \Delta_{AB}  ,$ or
$\Delta_{AB} \gg g_{AB}.$
For $\Delta_{AB}$ to be $\approx \chi_{AC}$, as desired,
$\chi_{AC}$ must be $\gg g_{AB} .$ This condition is achievable: in a scheme proposed in~\cite{Yan18}, two transmons achieve a capacitive-coupling $g_{AB}$ of 5 MHz. A dispersive shift $\chi_{AC}$ of a few hundred MHz is achievable, satisfying $\chi_{A C} \gg g_{AB}$~\cite{Ye24}. This outcome can result from custom coupling schemes such as the quarton coupler~\cite{Ye24}.

Having described our gate and its feasibility, we analyze the primary challenges to it: cavity loss and dephasing. After the clock populates the cavity, the population decays. The decay's extent depends on the gate-operation time, which dictates also the rotation angle in Eq.~\eqref{eq:XYgate}. The time-evolution unitary follows from the Hamiltonian
$H_{XY} = g_{AB}  \big( \sigma_+^{(A)}  \sigma_-^{(B)}  +  \hc  \big)
= \frac{g_{AB}}{2}  \sum_{a=x,y}  \sigma_a^{(A)}  \sigma_a^{(B)}$:
\begin{align}
    U_{AB}(t) = e^{-i  g_{{AB}}
    \left(  \sigma_x^{(A)} \otimes \sigma_x^{(B)}
             +  \sigma_y^{(A)} \otimes \sigma_y^{(B)}  \right)t / 2} \, .
    \label{eq:UXYt}
\end{align}
Let $\gamma_C$ denote the cavity's decay rate. The photon remains inside the cavity for a time $\approx 1/\gamma_C$. For the gate to impart a considerable rotation angle, $g_{AB}/\gamma_C$ must be at least $\approx 1$. A $\gamma_C$ comparable to $g_{AB} \approx5$ MHz (achievable according to the previous paragraph) is experimentally reasonable~\cite{Blais16, Krantz19}. One could optimize this gate's fidelity using simulations similar to those in App.~\ref{app_Fidelity}.

Like cavity loss, dephasing challenges our proposal. Dephasing can happen for two reasons. First, the cavity may not fully absorb the photon. This issue does not arise if the photon's bandwidth, $\Omega$, is much smaller than the cavity linewidth, $\gamma_C$. The parameter regime of, e.g., Sec.~\ref{sec:PhaseShift} meets this condition. Second, thermal photons in the coupled waveguide can enter the cavity and drive random gates that dephase $AB$’s state. Denote by $n_\text{th}$ the waveguide's average thermal-photon number.\footnote{
Let $T$ denote the waveguide's temperature, and let Boltzmann's constant equal 1. The average thermal-photon number is
$n_\text{th}= 1 / (e^{\omega_C / T}-1)$.}
Thermal photons impinge on the cavity at a rate $\gamma_C  \,  n_\text{th}$. A typical waveguide has a mode of frequency $\approx 6$ GHz and a temperature of 35 mK to 40 mK \cite{Blais16}. Under these conditions, $n_\text{th} \leq 10^{-3}$. Such a low thermal population can mitigate the transmon dephasing caused by stray photons.

\section{Outlook}
\label{sec:outlook}

We have proposed implementations of autonomous quantum gates for three experimental platforms: Rydberg atoms, trapped ions, and superconducting qubits. These gates could constitute building blocks for (partially or fully) autonomous quantum circuits. Greater autonomy could improve quantum devices' scalability, coherence times, and energy costs.  

Building quantum-autonomous computers would be challenging. However, our work shows that quantum-autonomous gates are feasible for current platforms, in principle. Scaling up to circuits will require architectural advances. Using neutral atoms, one must chain together optical fibers and lasers. Ions require complex trap geometries designable with finite-difference solvers.

In cavity QED and neutral-atom platforms, modularity could facilitate quantum-autonomous circuits: one would divide the circuit into modules. Each clock-tick photon would activate a module. A cascading mechanism would amplify the photon into multiple photons, which would activate different module components. To route the photons, one might draw inspiration from switches used in cavity QED \cite{Pechal16}.

Our results suggest several next steps. First, one could implement our gate proposals experimentally. After proof-of-principle experiments, one could incorporate our gates into quantum computers, helping free them from classical control. 
This incorporation could benefit quantum computation, e.g., as discussed at the end of Sec.~\ref{sec:Ultrafast}: quantum-autonomous PMLLs may stabilize ultrafast gates. Replacing Rydberg-blockade gates with ultrafast gates may reduce gate-operation times. Even our nonuniversal gate sets could enable quantum-autonomous Brownian circuits~\cite{Fisher23, Lashkari13, Shenker15, Zhou19, Xu19}, perhaps with U(1) symmetry~\cite{23_Fisher_Random}. 

Second, one could expand a platform's quantum-autonomous gates into a universal gate set. One could also identify mechanisms for chaining the gates together quantum-autonomously, as described in this section's second and third paragraphs. These expansions would enable universal quantum-autonomous circuits. 

Third, one could identify quantum-autonomous implementations of not only circuits, but quantum computations. Similar goals have been gaining traction at the intersection of atomic, molecular, and optical (AMO) physics with quantum error correction~\cite{Verstraete09, Leghtas13, Perez20, Gertler21, 21_Lebreuilly_Autonomous, Ghasemian23, Zapusek23, Wang23}.  Quantum thermodynamics can complement those fields' toolkits with AQM expertise~\cite{Mitchison19,MarinGuzman24}. Dissipative environments, studied across AMO physics and quantum thermodynamics, can assist~\cite{Verstraete09, Harrington22, Ghasemian23, Zapusek23, Wang23}. Quantum thermodynamics has already inspired a quantum-autonomous initialization of a superconducting qubit~\cite{Aamir25}. We envision, for each of several platforms, quantum-autonomous implementations of qubit initialization and reinitialization, circuits, and error correction.
When building a quantum computer, one could combine these quantum-autonomous components with classically controlled ones to optimize costs. 

Error mitigation, alluded to in the previous paragraph, offers a fourth opportunity. Quantum hardware suffers from coherent and incoherent errors. Coherent errors can arise due to slowly drifting parameters (e.g., a trapping laser's intensity). Gate sequences, such as spin echoes, can suppress these errors  \cite{Hahn50}. One can form such sequences from quantum-autonomous gates using autonomous quantum clocks (Sec. II B 1 and \cite{Meier26}). In contrast, incoherent errors stem from decoherence and so grow with a protocol's duration. Fast gates can suppress such errors; our ultrafast neutral-atom gate (Sec. \ref{sec:Ultrafast}) offers a model. Additionally, autonomous quantum error correction, mentioned in the previous paragraph, may suppress errors of multiple types.

Error correction may suit quantum autonomy due to its repetitiveness: one might parcel off repetitive tasks for quantum-autonomous components. For example, a quantum-autonomous circuit may facilitate measurements of a stabilizer generator. In cavity QED, single-flux-quantum logic may help embed such a circuit into a measurement process \cite{Liu23, Reitz26}.

Fifth, one might design quantum-autonomous alternatives to control techniques such as shortcuts to adiabaticity (STAs) \cite{Odelin19}. During STAs, time-varying control speeds up quantum evolutions. A global time-independent Hamiltonian, assisted by an autonomous quantum clock, can effect an arbitrary time-dependent local Hamiltonian \cite{Meier26, Page83}. Hence quantum machines might control their own STAs.

\begin{acknowledgments}
The authors thank Joseph Britton, Marko Cetina, Andrew Goffin, Marcus Huber, Mikhail Lukin, and Matteo Mitrano for valuable conversations. We also thank Sylvain de Léseleuc, Takuya Matsubara, and Kenji Ohmori for pertinent clarifications regarding ultrafast protocols on Rydberg atoms.
J.~A.~M.~G.~acknowledges support from NIST grant 70NANB21H055\_0. 
Y.-X.~W.~acknowledges support from a QuICS Hartree Postdoctoral Fellowship.
T.~M.~acknowledges support from the Harvard Quantum Initiative Postdoctoral Fellowship in Science and Engineering. 
P.~E.~acknowledges funding from the Austrian Federal Ministry of Education, Science and Research via the Austrian Research Promotion Agency (FFG) through the project FO999914033 (QUICHE) and from the European Research Council (Consolidator grant ``Cocoquest'’ 101043705).
S.~G.~acknowledges financial support from the European Research Council (Grant No. 101041744 ESQuAT) and from the Knut and Alice Wallenberg Foundation through the Wallenberg Centre for Quantum Technology (WACQT).
P.~E. and S.~G. are co-funded by the European Union (Quantum Flagship project ASPECTS, Grant Agreement No. 101080167). Views and opinions expressed are, however, those of the authors only and do not necessarily reflect those of the European Union, REA, or UKRI. Neither the European Union nor UKRI can beheld responsible for them.
N.~M.~L.~and N.~Y.~H. acknowledge support from the National Science Foundation (QLCI grant OMA-2120757). 

\end{acknowledgments}

\begin{appendices}

\onecolumngrid

\renewcommand{\thesection}{\Alph{section}}
\renewcommand{\thesubsection}{\Alph{section} \arabic{subsection}}
\renewcommand{\thesubsubsection}{\Alph{section} \arabic{subsection} \roman{subsubsection}}

\makeatletter\@addtoreset{equation}{section}
\def\theequation{\thesection\arabic{equation}}

\section{Quantum-autonomous Levine–Pichler gate}
\label{Levine_Pichler}

This appendix sketches a third quantum-autonomous controlled-phase gate for  {Rydberg} atoms. This gate complements the two in Sec.~\ref{RAAut}. First, we outline the protocol proposed by Levine and Pichler~\cite{Levine19}. It relies on global, rather than single-site, laser pulses. This relatively lax control seems promising for quantum autonomy. We argue that PMLLs can provide the required pulses and time the gate. Phase-locking the PMLLs stably poses a challenge, however. 

The Levine–Pichler gate leverages global laser pulses, instead of addressing atoms individually \cite{Levine19}. The protocol involves the computational-basis states
$\ket{0}=  \ket{5S_{1/2}, F{=}1, m_F {=} 0}$ and
$\ket{1}=  \ket{5S_{1/2}, F{=}2, m_F {=} 0}$, as well as the Rydberg state 
$\ket{\ryd}=  \ket{70S_{1/2}, m_J {=} -1/2}$. The protocol proceeds as follows:
\begin{enumerate}

 \item A bichromatic global laser pulse is resonant with the $\ket{1}$–$\ket{\ryd}$ transition. The atoms' dynamics depend on the initial state:
   \begin{itemize}
      
      \item If the atoms begin in $\ket{00}$, the pulse is off-resonant with all possible atomic transitions. The state remains constant.
      
      \item If the atoms begin in $\ket{01}$, the pulse drives the $\ket{01} \leftrightarrow \ket{0 \ryd}$ transition with a detuning $\Delta$ and Rabi frequency $\Omega$.
      
      \item If the atoms begin in $\ket{10}$, the pulse drives the $\ket{10} \leftrightarrow \ket{\ryd 0}$ transition with a detuning $\Delta$ and Rabi frequency $\Omega$.
      
      \item If the atoms begin in $\ket{11}$, the pulse drives the transition between $\ket{11}$ and the symmetric entangled state $\frac{1}{\sqrt{2}}(\ket{1 \ryd}+\ket{\ryd 1})$. The laser has a detuning $\Delta$, and the atoms oscillate with a Rabi frequency $\sqrt{2}  \,  \Omega$.
   
   \end{itemize}
   The pulse duration $\tau$ ensures that, if the atoms begin in $\ket{11}$, they undergo a complete Rabi cycle (return to $\ket{11}$). If the atoms begin in $\ket{01}$ or $\ket{10}$, they undergo only a partial oscillation.
   
   \item A second bichromatic pulse follows. It has the same duration as the first but has a phase offset: $\Omega \mapsto \Omega e^{i\xi}$. 
If the atoms began in $\ket{01}$ or $\ket{10}$, this pulse completes their Rabi oscillation. If the atoms began in $\ket{11}$, they undergo another full Rabi oscillation.
\end{enumerate}
The protocol maps each computational state to itself, while introducing a phase dependent on the detuning, $\Delta$. The choice $\Delta  = 0.377 \Omega$ implements a controlled-$Z$ gate.

PMLLs can quantum-autonomously emit pulses with the wavelengths and durations necessary to drive the Rydberg transition. One drives the $\ket{1}$–$\ket{\ryd}$ transition via a two-photon process, which requires two lasers. They need wavelengths of 420 nm and 1013 nm.\footnote{
The wavelengths differ from those mentioned in Sec.~\ref{RAAut} because the two protocols' intermediate states differ. Here, an atom jumps from $\ket{1}$ to $\ket{\ryd}$ via a sublevel of $\ket{6P_{3/2}}$. In the earlier protocol, an atom jumps via a different state $\ket{\rm i}$. }
One can tune PMLLs to emit at these wavelengths as described in Sec.~\ref{RAAut}: to achieve 420 nm, one can frequency-double a PMLL that emits at the commoner wavelength 840 nm. The second required wavelength, 1013 nm, is common among PMLLs. Furthermore, the required pulse lasts for $\tau = \frac{2\pi}{\sqrt{\Delta^2+2\Omega^2}} = 195$ ns. A PMLL can effect a pulse of this duration \cite{Liu15}. 

PMLLs can achieve not only the necessary wavelengths and durations, but also global pulses. A global pulse illuminates both atoms. Often, $\approx 5$ $\mu$m separate  {Rydberg} atoms~\cite{Saffman10}. Usual laser-beam waists are wider \cite{Silfvast04}.

Having argued that PMLLs can achieve three required specifications, we argue that a PMLL can time the Levine–Pichler gate. We argued in Sec.~\ref{sec:AutRydBl} that PMLLs can serve as autonomous quantum clocks. Hence an auxiliary laser could time and control the Levine–Pichler gate, as in Sec.~\ref{sec:AutRydBl}. To effect both pulses in the sequence, one would use two pairs of PMLLs, one pair per atom. Within each pair, the lasers must remain frequency-locked to each other with high stability, maintaining a constant phase shift between the two. If using PMLLs, one can control the pulse timing by controlling the length of the lasers' resonators. Yet small length changes and drifts can introduce large errors into the pulses' lengths \cite{PaschottaSync}. Hence PMLLs would render this gate challenging.

 \section{Time-dependent classical control in Rydberg-atom and superconducting-qubit proposals}
\label{app_Control}

Section~\ref{sec:traps} detailed a subtlety of our trapped-ion proposals: Paul traps are classical systems that consume and perform work, although not on the ions. The other platforms discussed (Rydberg atoms and superconducting qubits) require time-dependent classical control, too. Neither does these two platforms' control manipulate quantum-information-bearing DOFs, enacting gates. Hence our  {Rydberg}-atom and superconducting-qubit gates qualify as quantum-autonomous, according to the definition in~\cite{MarinGuzman24}. We support this claim for the two platforms consecutively. One might argue for a narrower definition of autonomous quantum machine, prohibiting time-dependent classical control of any DOF. This definition, however, would likely exclude nearly all quantum machines.

Classical time-dependent control holds  {Rydberg} atoms in place.  {If the atoms are neutral, t}hat control manifests as optical tweezers, highly focused laser beams \cite{Wu21}. In our proposals, optical tweezers only keep atoms fixed. The tweezers do not move atoms around to drive the gates, unlike in recent experiments on programmable neutral-atom simulators \cite{Bluvstein22}.

Classical control keeps superconducting devices at low temperatures. Superconducting devices operate inside dilution refrigerators whose workings include actively controlled elements. However, these elements do not manipulate information stored in the superconducting devices. Photons, emitted by an autonomous quantum clock, do. Similarly, a recent experiment demonstrated the quantum-autonomous initialization of a superconducting qubit in a dilution refrigerator \cite{Aamir25}.

\section{Second approach to halting quantum-autonomous controlled-$Z$ gate in circuit QED}
\label{app_SC_frac}

Section~\ref{sec:autcz} shows how to implement a controlled-$Z$ gate on flying (photonic) qubits in cavity QED, using an autonomous quantum clock. A conventional clock ticks repeatedly. In contrast, a simple gate implementation requires exactly one tick. Section~\ref{sec:autcz} presents one solution to this dilemma. We now sketch a more challenging approach.

In this approach, one engineers each clock tick to implement a fraction of the desired gate. Equation~\eqref{eq:XYgate} shows the desired gate; and Eq.~\eqref{eq:UXYt}, the gate implemented in a time $t$. Comparing the equations shows that, in a time $t$, the Bloch vector rotates through an angle $\theta = 2 g_{AB} t$. Let $\theta_0$ denote the desired angle. One may engineer the coupling $g_{AB}$ such that $\theta = \theta_0 / n$. The state will rotate through the desired angle after $n$ clock ticks.

One can limit the number of clock ticks by powering the clock with a finite energy source. An autonomous quantum clock's energy source consists of cold and hot baths (Fig.~\ref{fig:engine}). Once the clock depletes its energy source (once the baths come to the same temperature), the clock stops ticking.

This strategy endangers the gate's fidelity. In previous autonomous-quantum-clock analyses, researchers have assumed that the baths are infinitely large and Markovian~\cite{Erker17, Schwarzhans21, 23_Meier_Fundamental}. The finite-size baths we envision are non-Markovian. This property may alter the clock's reliability and so the gate's fidelity~\cite{23_Xuereb_Impact}. Furthermore, $n$ may vary across trials, as the clock's ticking is stochastic. Hence more infrastructure---perhaps a more complex clockwork~\cite{Schwarzhans21}---would be necessary to implement the gate reliably.

\section{Equations of motion for quantum-autonomous $Z$ gate on transmon}
\label{app_EOMs}

Section~\ref{sec:PhaseShift} introduced a quantum-autonomous protocol for implementing a $Z$ gate on a transmon. Here, we derive the effective equations of motion used to calculate the gate's fidelity. By ``effective,'' we mean that the equations reproduce the cavity-qubit dynamics described by the full master equation. First, we review the setup and protocol described in Sec.~\ref{sec:PhaseShift}. Then, we introduce the mathematical formalism and derive the effective dynamics, using quantum input-output theory~\cite{Gardiner85}.

Consider the following setup. A qubit corresponds to the Pauli-$z$ operator $\sigma_z$; and a cavity mode, to the raising operator $a^\dag$. A dispersive Hamiltonian $\frac{\chi}{2} \, a^\dag a \, \sigma_z$ couples the qubit to the cavity. The cavity couples also to an open waveguide, at a rate $\gamma$. 

The $Z$-gate protocol proceeds as follows. At time $t=0$, an autonomous quantum clock emits a single-photon Gaussian pulse of bandwidth $\Omega$. The pulse propagates through the waveguide to the cavity. The photon has a temporal profile
\begin{align}
   u (t) = \left ( \frac{\Omega^{2}}{2\pi}\right ) ^{1/4} 
   \exp \left  [ - \frac{\Omega ^{2}}{4} (t-t_{0}) ^{2} \right] .
\end{align}
The photon's frequency is detuned from the cavity mode by an amount $\Delta$. The photon reflects off the cavity, giving the transmon's state (expressed in terms of the computational basis) a relative phase. The photon thereafter travels to the opposite end of the waveguide.

To derive the effective equations of motion, we introduce notation for the input and output waveguide modes.
Let $a ^{\dag} _{\text{in}} (t) $ denote the input mode's raising operator, labeled by the time $t$. Let $ \ket{\text{vac}}$ denote the photonic DOF's vacuum state. The input photon is in the state
\begin{align}
   \label{eq_input_i}
   \ket{\Psi _{\text{in}}} 
   = \int ^{+\infty}_{-\infty}dt \: u (t)  \,  a ^{\dag} _{\text{in}} (t)  \, \ket{\text{vac}}.
\end{align}
Having introduced the input-mode notation, we introduce the output-mode notation. Let $a ^{\dag} _{\text{out}} (t) $ denote the output mode's raising operator. This field propagates oppositely the input field associated with $a ^{\dag} _{\text{in}} (t) $. 

Now, we formalize the modes', cavity's, and qubit's interdependence. 
Consider preparing the qubit in $\ket{\uparrow}$ or $\ket{\downarrow}$.
The cavity's state comes to depend on the qubit's state, via the dispersive coupling. The qubit-dependent cavity amplitudes $\bar{a} _{\uparrow/\downarrow} (t) $ satisfy the equations of motion
\begin{align}
\label{eq:eom.aup}
   & i \frac{d}{dt}  \,  \bar{a} _{\uparrow}(t)
   = \left [ \Delta + \frac{1}{2} (\chi - i \gamma)\right ] \bar{a} _{\uparrow}(t)  
   -i \sqrt{\gamma }  \, u (t) 
   \quad \text{and} \\ 
   \label{eq:eom.adown}
   & i \frac{d}{dt}  \,  \bar{a} _{\downarrow}(t)
   = \left [ \Delta - \frac{1}{2} (\chi + i \gamma)\right ] \bar{a} _{\downarrow}(t) 
   -i \sqrt{\gamma }  \, u (t) . 
\end{align}
These amplitudes inform the output amplitudes
$u _{\text{out}, \uparrow/\downarrow} (t)$,
which we can compute from the input-output relation
\begin{align}
   \label{eq:io}
   & u _{\text{out}, \uparrow/\downarrow} (t)  
   = u (t) + \sqrt{\gamma }  \;  \bar{a} _{\uparrow/\downarrow} (t)  .
\end{align}
All these ingredients determine the photonic state
\begin{align}
\label{eq:ph.wf}
   \ket{\Psi _{\uparrow/\downarrow} (t)} 
   = \int ^{+\infty}_{t} dt' \, u (t')  \,  a ^{\dag} _{\text{in}} (t') \ket{\text{vac}} 
   + \bar{a}   _{\uparrow/\downarrow} (t)  \,  a^{\dag} \ket{\text{vac}}
   + \int ^{t} _{-\infty} dt' \, 
   u _{\text{out}, \uparrow/\downarrow} (t')    \,
   a ^{\dag} _{\text{out}} (t') \ket{\text{vac}}.
\end{align}

Moreover, the ingredients above determine the qubit–cavity–waveguide system's state. Suppose the qubit begins in 
$ \cos \theta \ket{\uparrow} + e ^{i \phi} \sin \theta \ket{\downarrow}$. 
The input photon begins in $\ket{\Psi _{\text{in}}}$ [Eq.~\eqref{eq_input_i}]. Hence the qubit–cavity–waveguide system begins in
\begin{align}
   \ket{\Psi _{\text{tot}} (t{=}0) }
   = \left(  \cos \theta \ket{\uparrow} + e ^{i \phi} \sin \theta \ket{\downarrow}  \right) \otimes \ket{\Psi _{\text{in}}} .
\end{align}
The Hamiltonian commutes with the qubit's $\sigma _{z}$. Therefore, the qubit–cavity–waveguide system's state evolves as
\begin{align}
   \label{eq_Psi_tot}
   \ket{\Psi _{\text{tot}} (t)}= \cos \theta \ket{\uparrow} \ket{\Psi _{\uparrow} (t)} + e ^{i \phi} \sin \theta \ket{\downarrow} \ket{\Psi _{\downarrow} (t)} .
\end{align}

We can now compute the time-dependent qubit coherence function, defined as follows. Denote the qubit's time-$t$ density operator by $\rho_Q(t)$. Consider the matrix that represents $\rho_Q(t)$ relative to the computational basis, 
$\Set{ \ket{\uparrow}, \ket{\downarrow} }$. The top off-diagonal element,
$\bra{\uparrow}  \rho _Q (t) \ket{\downarrow}$,
quantifies the state's coherence. To calculate this matrix element, we trace out the photon ($P$) from the whole-system state $ \ket{\Psi _{\text{tot}} (t)}$ [Eq.~\eqref{eq_Psi_tot}]:
\begin{align}
   \bra{\uparrow}  \rho _Q (t) \ket{\downarrow} 
   & = \text{Tr}  _P \LParen
   \langle {\uparrow}  
   \ketbra{\Psi _{\text{tot}} (t) } {\Psi _{\text{tot}} (t) }   
   {\downarrow} \rangle \RParen
   = e ^{-i \phi} \cos \theta \sin \theta \, 
   \braket{\Psi _{\downarrow} (t)} { \Psi _{\uparrow} (t)}  \\
   & = \bra{\uparrow}  \rho _Q (-\infty) \ket{\downarrow}  
   \braket{\Psi _{\downarrow} (t)} { \Psi _{\uparrow} (t)} .
\end{align}
The final factor---the photonic-state overlap---follows from Eq.~\eqref{eq:ph.wf}:
\begin{align}
   \braket{\Psi _{\downarrow} (t)} { \Psi _{\uparrow} (t)}
   = \int ^{+\infty}_{t} dt' \, |u (t') |^{2} 
   +\bar{a} _{\uparrow}(t)  \,
   \bar{a} ^{*} _{\downarrow}(t) 
   +  \int ^{t} _{-\infty} dt' \, u _{\text{out}, \uparrow} (t') \,
   u ^{*} _{\text{out}, \downarrow} (t') .
\end{align}
Into this equation, we substitute the input-output relation~\eqref{eq:io}:
\begin{align}
   & \braket{\Psi _{\downarrow} (t)} { \Psi _{\uparrow} (t)}
   =  \int ^{+\infty}_{t} dt' \, |u (t') |^{2} 
   +\bar{a} _{\uparrow} (t)  \,
   \bar{a} ^{*} _{\downarrow} (t)+  \int ^{t} _{-\infty} dt' \, 
   [u (t') + \sqrt{\gamma } \:  \bar{a} _{\uparrow} (t') ] 
   [u ^{*} (t') + \sqrt{\gamma } \:  \bar{a}  ^{*}_{\downarrow} (t') ]
   \nonumber \\
   \label{eq:qb.coh.int}
   & =  \int ^{+\infty}_{-\infty} dt' \, |u (t') |^{2} +\bar{a} _{\uparrow} (t)
   \bar{a} ^{*} _{\downarrow} (t) 
   +  \int ^{t} _{-\infty} dt' 
   \left\{ \sqrt{\gamma }  \:
   \left [\bar{a} _{\uparrow} (t')  \,  u ^{*} (t') 
     + u (t')  \,  \bar{a}  ^{*}_{\downarrow} (t')   \right] 
   + \gamma \bar{a} _{\uparrow} (t')  \,   \bar{a}  ^{*}_{\downarrow} (t')  \right\} 
   \nonumber \\
   & =  1 +\bar{a} _{\uparrow} (t)  \,  \bar{a} ^{*} _{\downarrow} (t) 
   +  \int ^{t} _{-\infty} dt' 
   \left\{ \sqrt{\gamma }  \:
   \left[  \bar{a} _{\uparrow} (t')  \,  u ^{*} (t') 
            + u (t')  \,  \bar{a}  ^{*}_{\downarrow} (t')  \right] 
   + \gamma \bar{a} _{\uparrow} (t')  \,  \bar{a}  ^{*}_{\downarrow} (t')  \right\} .
\end{align}
To simplify the integral, we use the cavity-mode equations of motion, Eqs.~\eqref{eq:eom.aup} and~\eqref{eq:eom.adown}. They imply
\begin{align}
   & \bar{a}  ^{*}_{\downarrow}(t)  \, \frac{d}{dt}  \,  \bar{a} _{\uparrow}(t) 
   = \left [ - i \Delta - \frac{i\chi }{2} -\frac{\gamma }{2} \right ] 
   \bar{a}  ^{*}_{\downarrow}(t)   \, \bar{a} _{\uparrow} (t)  
   - \sqrt{\gamma }  \: u (t)  \, \bar{a}  ^{*}_{\downarrow}(t)  
   \quad \text{and} \\ 
   & \bar{a} _{\uparrow}(t)   \,  \frac{d}{dt}  \, \bar{a}  ^{*}_{\downarrow} (t) 
   = \left [  i \Delta - \frac{i\chi }{2} 
   -\frac{\gamma }{2} \right ] 
   \bar{a}  ^{*}_{\downarrow}(t)   \, \bar{a} _{\uparrow} (t)  
   - \sqrt{\gamma }  \: u (t)  \, \bar{a} _{\uparrow} (t)  \,  .
\end{align}
Summing these two equations yields
\begin{align}
   & \frac{d}{dt}  \, \left[ \bar{a}  ^{*}_{\downarrow}(t)   \,  \bar{a} _{\uparrow}(t)   \right] 
   = (- i\chi -\gamma )  \, \bar{a}  ^{*}_{\downarrow}(t)   \,  \bar{a} _{\uparrow} (t)  
   - \sqrt{\gamma }  \: 
   \left[ u (t)  \, \bar{a}  ^{*}_{\downarrow}(t)  
          +  u (t)  \, \bar{a} _{\uparrow} \right(t) ]  \\
   \label{eq:cav.prod.timederiv}
    \Rightarrow \quad&
   \frac{d}{dt}  \, \left[ \bar{a}  ^{*}_{\downarrow}(t)   \,  \bar{a} _{\uparrow}(t)   \right] 
   + \gamma \bar{a}  ^{*}_{\downarrow}(t)   \,   \bar{a} _{\uparrow} (t) 
   + \sqrt{\gamma }  \: 
   \left[ u (t)  \, \bar{a}  ^{*}_{\downarrow}(t)  +u (t)  \, \bar{a} _{\uparrow}(t)  \right]  
   = - i\chi \bar{a}  ^{*}_{\downarrow}(t)   \, \bar{a} _{\uparrow} (t)  \,  .
\end{align}
Integrating both sides of Eq.~\eqref{eq:cav.prod.timederiv}, we recover the integral in Eq.~\eqref{eq:qb.coh.int}:
\begin{align}
   & \int ^{t} _{-\infty} dt'  
   \left \{ \frac{d}{dt'}  \left[ \bar{a}  ^{*}_{\downarrow}(t') \bar{a} _{\uparrow}(t')   \right] 
   + \gamma  \,  \bar{a}  ^{*}_{\downarrow}(t')   \,  \bar{a} _{\uparrow}(t')  
   + \sqrt{\gamma }  \: 
   \left[ u (t')  \,  \bar{a}  ^{*}_{\downarrow}(t')  
           +u (t')  \,  \bar{a} _{\uparrow}(t')    \right]  \right \} 
   = - i\chi \int ^{t} _{-\infty} dt'  \,
   \bar{a}^{*}_{\downarrow} (t')  \,  \bar{a} _{\uparrow}(t')  \\
   & = \bar{a} _{\uparrow} (t)  \,  \bar{a} ^{*} _{\downarrow} (t) 
   +  \int ^{t} _{-\infty} dt' 
   \left\{ \sqrt{\gamma } \:
   \left[  \bar{a} _{\uparrow} (t')  \,  u ^{*} (t') 
           + u (t')\bar{a}  ^{*}_{\downarrow} (t')  \right] 
   + \gamma \bar{a} _{\uparrow} (t')  \,   \bar{a}  ^{*}_{\downarrow} (t')  \right\} .
\end{align}
Substituting into Eq.~\eqref{eq:qb.coh.int}, we obtain the simplified qubit coherence function:
\begin{align}
   \frac{\bra{\uparrow}  \rho _Q (t) 
   \ket{\downarrow}}{\bra{\uparrow}  \rho _Q (-\infty) \ket{\downarrow} } 
   = \braket{\Psi _{\downarrow} (t)} { \Psi _{\uparrow} (t)} 
   = 1 - i\chi \int ^{t} _{-\infty} dt' \:
   \bar{a}  ^{*}_{\downarrow} (t')  \, \bar{a} _{\uparrow}(t')  .
   \label{eq_Coher_fxn_1}
\end{align}

The change in the qubit coherence function, 
$\langle \sigma _{-} (t) \rangle -  \langle \sigma _{-} (0) \rangle$, equals the following quantity. Suppose the input field began in a coherent state whose average photon number equaled $1$. Consider the change, from time 0 to time $t$, in the coherence function's logarithm. This hypothetical change equals the true change,
$\langle \sigma _{-} (t) \rangle -  \langle \sigma _{-} (0) \rangle$.
This equality is no coincidence: using third quantization~\cite{McDonald23}, one can show the following about a dispersively coupled qubit-cavity-waveguide system. The change in the coherence function, given a single-photon input, equals the change in the coherence function's logarithm, given a coherent-state input. This equality stems from
the following mathematical property: the coherence, given a single-photon input, is the generating function for the coherence given a coherent (photonic) state. This correspondence characterizes general qubit–boson systems governed by dispersive Hamiltonians, which can be solved analytically.\footnote{
More generally, the correspondence characterizes every qubit–boson system Hamiltonian that (i) conserves the qubit's energy and (ii) is quadratic in the bosonic raising and lowering operators.}

Let us summarize the derivations above. We can compute the qubit coherence dynamics via the equations of motion
\begin{align}
   \label{eq:EOMs}
   \begin{cases}
i\frac{d}{dt} \langle \sigma _{-} (t) \rangle 
   =\chi   \langle \sigma _{-} (-\infty) \rangle  \, 
   \bar{a} _{\uparrow}(t)   \,  \bar{a} ^{*} _{\downarrow}(t) \\
   i \frac{d}{dt} \bar{a} _{\uparrow}(t)  
   = \left [ \Delta + \frac{1}{2} (\chi - i \gamma)\right ] \bar{a} _{\uparrow}(t)   
   -i \sqrt{\gamma }  \: u (t)  \\ 
   i \frac{d}{dt} \bar{a} _{\downarrow}(t)  
   = \left [ \Delta - \frac{1}{2} (\chi + i \gamma)\right ] \bar{a} _{\downarrow}(t)   
   -i \sqrt{\gamma }  \: u (t)
   \end{cases} 
\end{align}

Let us identify the parameter regime that supports a high-fidelity qubit gate. We wish to minimize the distortions in the outgoing photon: they can result from entanglement between the qubit and the output photon. Such entanglement diminishes the qubit's coherence. Hence we focus on the regime in which the waveguide interacts with the cavity over a time much shorter than the pulse duration. In this narrow-bandwidth regime, $\Omega \ll \gamma $. The input pulse scatters off the cavity before the interaction affects the photon-pulse shape. Therefore, we can adiabatically approximate the cavity amplitudes:
\begin{align}
   |\partial _{t}  \, \bar{a} _{\uparrow,\downarrow}(t) | 
   \ll \gamma |\bar{a} _{\uparrow,\downarrow}(t)  | 
   \quad \Rightarrow  \quad
   \begin{cases}
   \bar{a} _{\uparrow} (t) 
   \simeq  \frac{i \sqrt{\gamma } }{ \Delta + \frac{\chi - i \gamma}{2}  } \,  u (t) \\
   \bar{a} _{\downarrow} (t) \simeq \frac{i \sqrt{\gamma }}{\Delta - \frac{\chi + i \gamma}{2} }  u (t) .
   \end{cases}
\end{align}
Let us substitute these approximate qubit-dependent cavity amplitudes into the qubit coherence function~\eqref{eq_Coher_fxn_1}:
\begin{align}
   \frac{\bra{\uparrow}  \rho _Q (t) \ket{\downarrow}}{\bra{\uparrow}  
   \rho _Q (-\infty) \ket{\downarrow} } 
   =  1 - i\chi \int ^{t} _{-\infty} dt' \;  
   \bar{a}  ^{*}_{\downarrow} (t') \,  \bar{a} _{\uparrow}(t')  
   \simeq 1 - \frac{i \chi \gamma }{\left ( \Delta + \frac{\chi - i \gamma}{2} \right )\left ( \Delta - \frac{\chi - i \gamma}{2}  \right )} 
   \int ^{t} _{-\infty} dt' \,  |u (t') |^2  .
\end{align}
At long times ($t \gg \Omega ^{-1}$), the qubit coherence function approximates to
\begin{align}
\lim _{t\to \infty}\frac{\bra{\uparrow}  \rho _Q (t) \ket{\downarrow}}{\bra{\uparrow}  \rho _Q (-\infty) \ket{\downarrow} } 
\simeq 1 - \frac{i \chi \gamma }{\left ( \Delta + \frac{\chi - i \gamma}{2} \right )\left ( \Delta - \frac{\chi - i \gamma}{2}  \right )} .
\end{align}
The detuning value $\Delta = \frac{\chi }{2}$ maximizes the left-hand side's magnitude, minimizing the qubit dephasing. Under this condition, the coherence function satisfies the approximation
\begin{align}
   \lim _{t \to \infty}
   \frac{\bra{\uparrow}  \rho _Q (t) \ket{\downarrow} }{
   \bra{\uparrow} \rho _Q (-\infty) \ket{\downarrow} } 
   \simeq 1 - \frac{2\chi}{\chi - \frac{ i \gamma}{2} } 
   = \frac{\frac{  \gamma}{2} - i\chi}{\frac{  \gamma}{2} + i\chi}
   = \exp \left( -2i \arctan \frac{2\chi }{\gamma} \right) . 
   \label{eq_Coher_fxn_2}
\end{align}
By the left-hand side's definition, the qubit's Bloch vector rotates about the $z$-axis through an angle
\begin{align}
\phi = 2\arctan (2\chi /\gamma)  .
\end{align}
We have recovered Eq.~\eqref{eq:Shift}.

\section{Fidelity of second quantum-autonomous $Z$ gate for transmon}
\label{app_Fidelity}

Section~\ref{sec:PhaseShift} introduced a quantum-autonomous $Z$ gate implemented on a transmon via a dispersive coupling. Here, we calculate the gate's fidelity. We numerically show that the qubit reaches a steady state in a time $\approx 200 / \gamma$, as claimed at the end of Sec.~\ref{sec:PhaseShift}. Finally, we identify the optimal parameter regime.

We calculate the fidelity as follows. Consider initializing the transmon in the state
$\ket{\psi}=a\ket{0}+b\ket{1}$, wherein $a, b \in \mathbb{C}$ and $| a |^2 + | b |^2 = 1$. Suppose we wish to rotate the Bloch vector about the $z$-axis through an angle $\phi$. Ideally, the state evolves to
$\ket{\psi_\phi} = a\ket{0}+be^{i\phi}\ket{1}$. The actual final state, $\rho$, may suffer from imperfections. Define as $\rho_{jk} \coloneqq \bra{j} \rho \ket{k}$ the $(j, k)$ element of the matrix that represents $\rho$ relative to the computational basis, wherein
$j, k \in \{0, 1\}$. We aim to compute the fidelity between 
$\rho$ and $\ketbra{\psi_\phi}{\psi_\phi}$. Since the latter state is pure, the fidelity reduces to~\cite{NielsenC10}
\begin{align}
   \label{eq_F_help1}
    &F \left( \rho, \ketbra{\psi_\phi}{\psi_\phi} \right) 
    =\bra{\psi_\phi}\rho\ket{\psi_\phi}
    =|a|^2\bra{0}\rho_{00}\ket{0}+a^*be^{i\phi}\bra{0}\rho_{01}\ket{1}
    +ab^*e^{-i\phi}\bra{1}\rho_{10}\ket{0}+|b|^2\bra{1}\rho_{11}\ket{1}.
\end{align}
The penultimate term depends on $\rho_{10}=\expval{\sigma_-}$. 
Since the qubit couples to the cavity dispersively, the gate preserves the initial state's populations: 
$\rho_{00} = |a|^2$, and $\rho_{11} = |b|^2$. Equation~\eqref{eq_F_help1} simplifies to
\begin{align}
   \label{eq_F_help2}
    F \left( \rho, \ketbra{\psi_\phi}{\psi_\phi} \right) 
    =|a|^4+|b|^4+2\text{Re}  \left(  ab^*e^{-i\phi}\langle\sigma_-\rangle  \right) .
\end{align}

\begin{figure}[h]
    \includegraphics[width=0.83\linewidth]{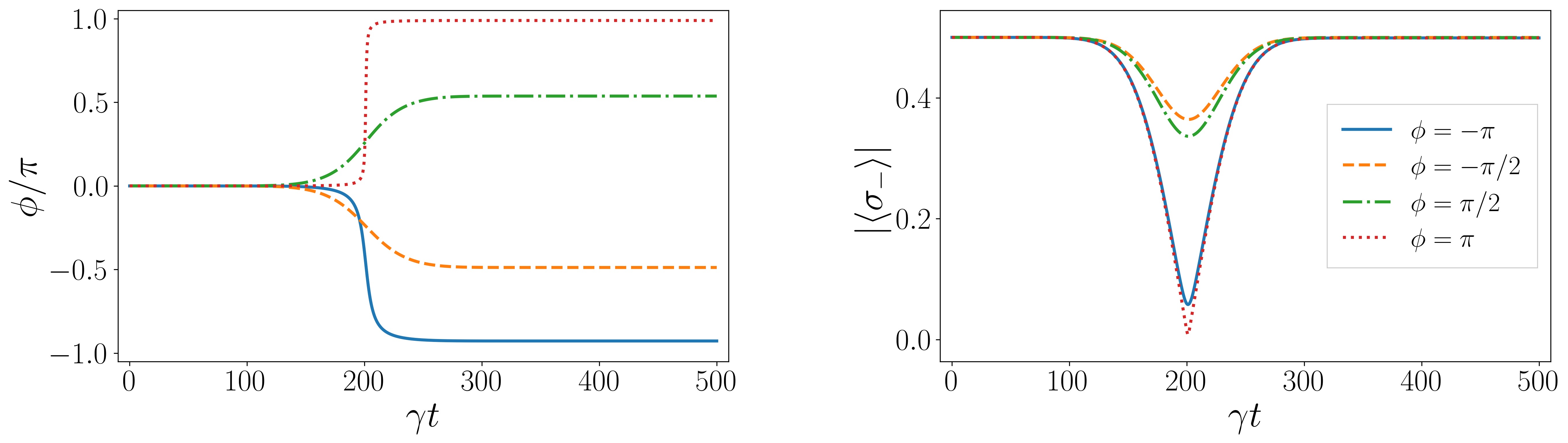}
        \caption{Time-dependent effects of a photon on a superconducting qubit coupled dispersively to a cavity.
        The left-hand plot shows the angle through which the qubit's Bloch vector has rotated about the $z$-axis by the time $t$. The right-hand plot quantifies the state's coherence. Each $x$-axis shows time measured in units of $1/\gamma$, the cavity's inverse lifetime. We formed each curve using the parameters that enable the best-fidelity approximation to a rotation through the corresponding angle $\phi$.}
        \label{fig:ZGate}
\end{figure}

Having calculated the fidelity, we numerically illustrate the phase and coherence achievable throughout the gate protocol. We solve Eqs.~\eqref{eq:EOMs} for the coherence function $\expval{\sigma_-}  \equiv  |\langle{\sigma_-}\rangle|e^{i\phi}$. The coherence magnitude $|\langle{\sigma_-}\rangle|$ and phase $\phi$ depend on the time $t$ for which the gate has operated. Figure~\ref{fig:ZGate} depicts these time dependences. $|\langle{\sigma_-}\rangle|$ and $\phi$ stabilize over a time scale $\approx 200/\gamma$.

\begin{figure}[h!]
  \includegraphics[width=0.5\linewidth]{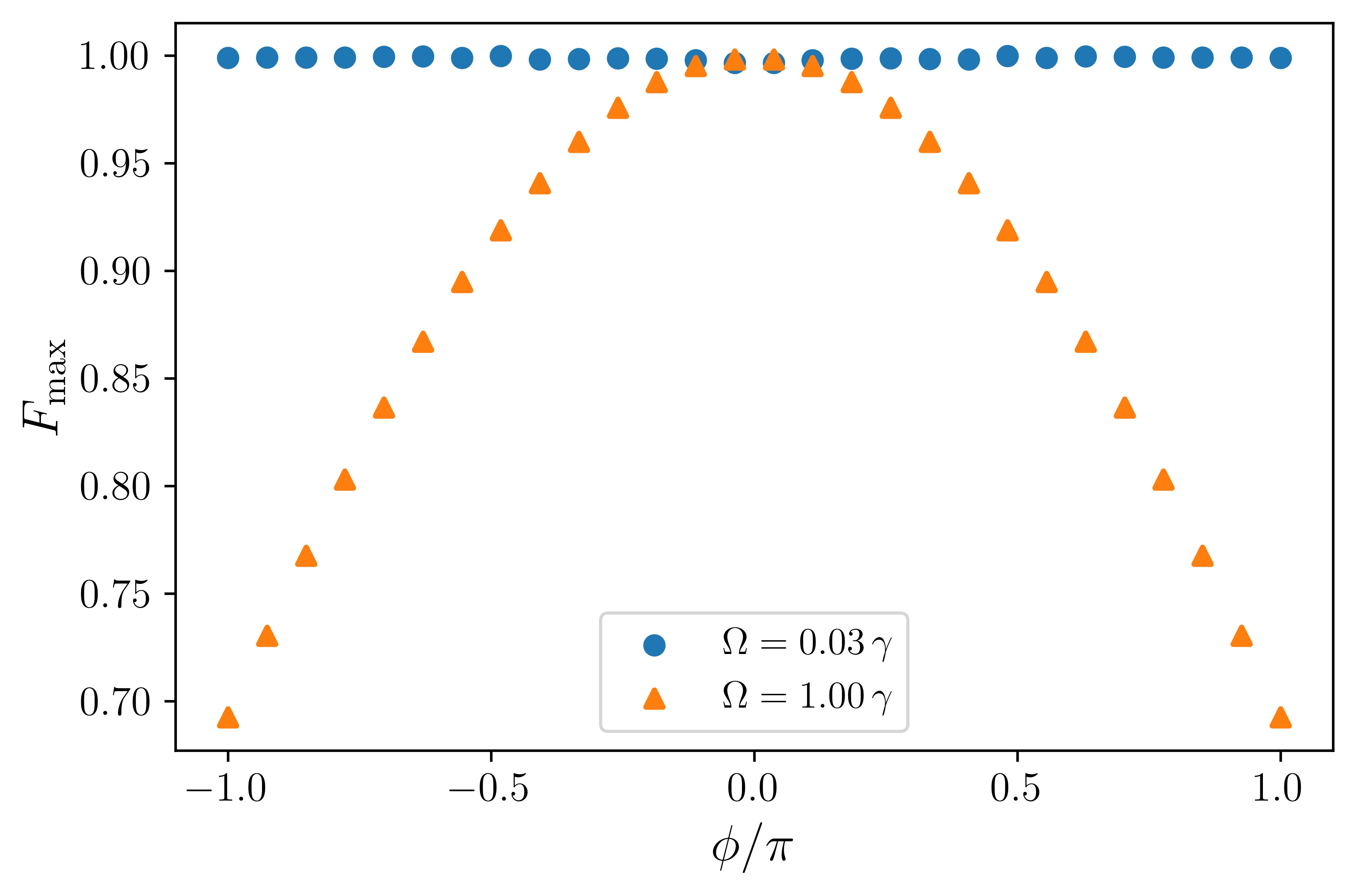}
        \caption{Fidelity achievable with the main text's second quantum-autonomous superconducting-qubit $Z$ gate. In the narrow-bandwidth, strong-coupling regime $\Omega\ll\gamma\ll\chi$ (blue disks), the fidelity is near-perfect. If $\Omega$ is tuned out of this regime (orange triangles), the fidelity worsens as the rotation-angle magnitude $| \phi |$ grows. }
        \label{fig:FidMax}
\end{figure}

We identified the optimal parameter regime as follows. First, we fixed the cavity lifetime $\gamma$, the unit of inverse time. $\gamma$ must far exceed the photon bandwidth, $\gamma \gg \Omega$, so we set $\Omega = 0.03\gamma$. Fixing the photon-cavity detuning at $\Delta=\chi/2$, we swept $\chi$ over $[0,30\gamma]$. Figure \ref{fig:FidMax} shows the maximum attainable fidelity as a function of the target angle, $\phi$. We define this maximum fidelity as the greatest overlap $F( \rho, \ketbra{\psi_\phi}{\psi_\phi})$ between the evolved qubit state $\rho$ and the target state $\ket{\psi_\phi}$, optimized over all accessible parameter sets. At small angles $\phi \gtrsim 0$, the fidelity maximizes when $\chi\approx0$. At large angles $\phi \approx \pi$, the fidelity maximizes when $\gamma\ll\chi$. For example, the protocol can realize a $\pi/2$-rotation with a fidelity of 0.9989 and a coherence of 0.4997 (right-hand plot of Fig. \ref{fig:ZGate}).

\end{appendices}
\newpage

%
\bibliographystyle{h-physrev}
\bibliography{AQGs}

@article{MarinGuzman24,
doi = {10.1088/1361-6633/ad8803},
year = {2024},
month = {nov},
publisher = {IOP Publishing},
volume = {87},
number = {12},
pages = {122001},
author = {Marín Guzmán, José Antonio and Erker, Paul and Gasparinetti, Simone and Huber, Marcus and Yunger Halpern, Nicole},
title = {Key issues review: useful autonomous quantum machines},
journal = {Rep. Prog. Phys.}
}

@article{Sorensen99,
  title = {Quantum Computation with Ions in Thermal Motion},
  author = {S\o{}rensen, Anders and M\o{}lmer, Klaus},
  journal = {Phys. Rev. Lett.},
  volume = {82},
  issue = {9},
  pages = {1971--1974},
  numpages = {0},
  year = {1999},
  month = {Mar},
  publisher = {American Physical Society},
  doi = {10.1103/PhysRevLett.82.1971}
}

@article{Auffeves22,
  title = {Quantum Technologies Need a Quantum Energy Initiative},
  author = {Auff\`eves, Alexia},
  journal = {PRX Quantum},
  volume = {3},
  issue = {2},
  pages = {020101},
  numpages = {12},
  year = {2022},
  month = {Jun},
  publisher = {American Physical Society},
  doi = {10.1103/PRXQuantum.3.020101}
}

@Article{Maslennikov19,
author={Maslennikov, Gleb
and Ding, Shiqian
and Habl{\"u}tzel, Roland
and Gan, Jaren
and Roulet, Alexandre
and Nimmrichter, Stefan
and Dai, Jibo
and Scarani, Valerio
and Matsukevich, Dzmitry},
title={Quantum absorption refrigerator with trapped ions},
journal={Nat. Commun.},
year={2019},
month={Jan},
day={14},
volume={10},
number={1},
pages={202},
issn={2041-1723},
doi={10.1038/s41467-018-08090-0}
}

@Article{Sackett00,
author={Sackett, C. A.
and Kielpinski, D.
and King, B. E.
and Langer, C.
and Meyer, V.
and Myatt, C. J.
and Rowe, M.
and Turchette, Q. A.
and Itano, W. M.
and Wineland, D. J.
and Monroe, C.},
title={Experimental entanglement of four particles},
journal={Nature},
year={2000},
month={Mar},
day={01},
volume={404},
number={6775},
pages={256-259},
issn={1476-4687},
doi={10.1038/35005011}
}

@Article{Aamir25,
author={Aamir, Mohammed Ali
and Jamet Suria, Paul
and Mar{\'i}n Guzm{\'a}n, Jos{\'e} Antonio
and Castillo-Moreno, Claudia
and Epstein, Jeffrey M.
and Yunger Halpern, Nicole
and Gasparinetti, Simone},
title={Thermally driven quantum refrigerator autonomously resets a superconducting qubit},
journal={Nat. Phys.},
year={2025},
month={Feb},
day={01},
volume={21},
number={2},
pages={318-323},
issn={1745-2481},
doi={10.1038/s41567-024-02708-5}
}

@Article{Bluvstein22,
author={Bluvstein, Dolev
and Levine, Harry
and Semeghini, Giulia
and Wang, Tout T.
and Ebadi, Sepehr
and Kalinowski, Marcin
and Keesling, Alexander
and Maskara, Nishad
and Pichler, Hannes
and Greiner, Markus
and Vuleti{\'{c}}, Vladan
and Lukin, Mikhail D.},
title={A quantum processor based on coherent transport of entangled atom arrays},
journal={Nature},
year={2022},
month={Apr},
day={01},
volume={604},
number={7906},
pages={451-456},
issn={1476-4687},
doi={10.1038/s41586-022-04592-6}
}

@article{Mitchison19,
author = {Mark T. Mitchison},
title = {Quantum thermal absorption machines: refrigerators, engines and clocks},
journal = {Contemp. Phys.},
volume = {60},
number = {2},
pages = {164-187},
year = {2019},
publisher = {Taylor & Francis},
doi = {10.1080/00107514.2019.1631555}

}

@article{Woods23,
  title = {Autonomous Quantum Devices: When Are They Realizable without Additional Thermodynamic Costs?},
  author = {Woods, Mischa P. and Horodecki, Micha\l{}},
  journal = {Phys. Rev. X},
  volume = {13},
  issue = {1},
  pages = {011016},
  numpages = {31},
  year = {2023},
  month = {Feb},
  publisher = {American Physical Society},
  doi = {10.1103/PhysRevX.13.011016}
}

@article{Schwarzhans21,
  title = {Autonomous temporal probability concentration: Clockworks and the second Law of thermodynamics},
  author = {Schwarzhans, Emanuel and Lock, Maximilian P. E. and Erker, Paul and Friis, Nicolai and Huber, Marcus},
  journal = {Phys. Rev. X},
  volume = {11},
  issue = {1},
  pages = {011046},
  numpages = {22},
  year = {2021},
  month = {Mar},
  publisher = {American Physical Society},
  doi = {10.1103/PhysRevX.11.011046}
}

@article{Erker17,
  title = {Autonomous quantum clocks: Does thermodynamics limit our ability to measure time?},
  author = {Erker, Paul and Mitchison, Mark T. and Silva, Ralph and Woods, Mischa P. and Brunner, Nicolas and Huber, Marcus},
  journal = {Phys. Rev. X},
  volume = {7},
  issue = {3},
  pages = {031022},
  numpages = {12},
  year = {2017},
  month = {Aug},
  publisher = {American Physical Society},
  doi = {10.1103/PhysRevX.7.031022}
}

@article{Meier26,
doi = {10.1088/1361-6633/ae7b82},
url = {https://doi.org/10.1088/1361-6633/ae7b82},
year = {2026},
month = {jul},
publisher = {IOP Publishing},
volume = {89},
number = {7},
pages = {077601},
author = {Meier, Florian and Huber, Marcus and Erker, Paul and Xuereb, Jake},
title = {Autonomous quantum processing unit: an autonomous thermal computing machine &amp; its physical limitations},
journal = {Rep. Prog. Phys.}
}

@ARTICLE{24_Woods_Quantum,
       author = {{Woods}, Mischa P.},
        title = "{Quantum Frequential Computing: a quadratic runtime advantage for all computations}",
      journal = {arXiv e-prints},
         year = 2024,
        month = mar,
          eid = {arXiv:2403.02389},
        pages = {arXiv:2403.02389},
          doi = {10.48550/arXiv.2403.02389},
archivePrefix = {arXiv},
       eprint = {2403.02389},
 primaryClass = {quant-ph},
       adsurl = {https://ui.adsabs.harvard.edu/abs/2024arXiv240302389W}
}

@article{Jaksch00,
  title = {Fast Quantum Gates for Neutral Atoms},
  author = {Jaksch, D. and Cirac, J. I. and Zoller, P. and Rolston, S. L. and C\^ot\'e, R. and Lukin, M. D.},
  journal = {Phys. Rev. Lett.},
  volume = {85},
  issue = {10},
  pages = {2208--2211},
  numpages = {0},
  year = {2000},
  month = {Sep},
  publisher = {American Physical Society},
  doi = {10.1103/PhysRevLett.85.2208}
}

@article{Wang23,
  title = {Quantum nonreciprocal interactions via dissipative gauge symmetry},
  author = {Wang, Yu-Xin and Wang, Chen and Clerk, Aashish A.},
  journal = {PRX Quantum},
  volume = {4},
  issue = {1},
  pages = {010306},
  numpages = {28},
  year = {2023},
  month = {Jan},
  publisher = {American Physical Society},
  doi = {10.1103/PRXQuantum.4.010306}
}

@article{Guerreiro21,
  title = {Molecular machines for quantum error correction},
  author = {Guerreiro, Thiago},
  journal = {PRX Quantum},
  volume = {2},
  issue = {3},
  pages = {030336},
  numpages = {16},
  year = {2021},
  month = {Aug},
  publisher = {American Physical Society},
  doi = {10.1103/PRXQuantum.2.030336}
}

@ARTICLE{Sundelin24,
       author = {{Sundelin}, Simon and {Aamir}, Mohammed Ali and {Manish Kulkarni}, Vyom and {Castillo-Moreno}, Claudia and {Gasparinetti}, Simone},
        title = "{Quantum refrigeration powered by noise in a superconducting circuit}",
      journal = {arXiv e-prints},
         year = 2024,
        month = mar,
          eid = {arXiv:2403.03373},
        pages = {arXiv:2403.03373},
          doi = {10.48550/arXiv.2403.03373},
archivePrefix = {arXiv},
       eprint = {2403.03373},
 primaryClass = {quant-ph},
       adsurl = {https://ui.adsabs.harvard.edu/abs/2024arXiv240303373S}
}

@Article{Harrington22,
author={Harrington, Patrick M.
and Mueller, Erich J.
and Murch, Kater W.},
title={Engineered dissipation for quantum information science},
journal={Nat. Rev. Phys.},
year={2022},
month={Oct},
day={01},
volume={4},
number={10},
pages={660-671},
issn={2522-5820},
doi={10.1038/s42254-022-00494-8}
}

@article{Xu19,
  title = {Locality, Quantum Fluctuations, and Scrambling},
  author = {Xu, Shenglong and Swingle, Brian},
  journal = {Phys. Rev. X},
  volume = {9},
  issue = {3},
  pages = {031048},
  numpages = {21},
  year = {2019},
  month = {Sep},
  publisher = {American Physical Society},
  doi = {10.1103/PhysRevX.9.031048}
}

@article{Li17,
  title = {Realization of Translational Symmetry in Trapped Cold Ion Rings},
  author = {Li, Hao-Kun and Urban, Erik and Noel, Crystal and Chuang, Alexander and Xia, Yang and Ransford, Anthony and Hemmerling, Boerge and Wang, Yuan and Li, Tongcang and H\"affner, Hartmut and Zhang, Xiang},
  journal = {Phys. Rev. Lett.},
  volume = {118},
  issue = {5},
  pages = {053001},
  numpages = {5},
  year = {2017},
  month = {Jan},
  publisher = {American Physical Society},
  doi = {10.1103/PhysRevLett.118.053001}
}

@article{Wang15,
doi = {10.1088/0953-4075/48/20/205002},
year = {2015},
month = {sep},
publisher = {IOP Publishing},
volume = {48},
number = {20},
pages = {205002},
author = {Po-Jen Wang and Tongcang Li and C Noel and A Chuang and Xiang Zhang and H Häffner},
title = {Surface traps for freely rotating ion ring crystals},
journal = {J. Phys. B}
}

@article{Urban19,
  title = {Coherent Control of the Rotational Degree of Freedom of a Two-Ion {Coulomb} Crystal},
  author = {Urban, Erik and Glikin, Neil and Mouradian, Sara and Krimmel, Kai and Hemmerling, Boerge and Haeffner, Hartmut},
  journal = {Phys. Rev. Lett.},
  volume = {123},
  issue = {13},
  pages = {133202},
  numpages = {5},
  year = {2019},
  month = {Sep},
  publisher = {American Physical Society},
  doi = {10.1103/PhysRevLett.123.133202}
}

@misc{Preskill23,
  author = {Preskill, John},
  journal = {Quantum Frontiers},
  title = {Crossing the quantum chasm: {From NISQ} to fault tolerance},
  year = {2023},
  month = {Dec},
  howpublished = {Quantum Frontiers},
  note = {{Available online at \url{https://quantumfrontiers.com/2023/12/09/crossing-the-quantum-chasm-from-nisq-to-fault-tolerance/}}}
}

@Article{Shenker15,
author={Shenker, Stephen H.
and Stanford, Douglas},
title={Stringy effects in scrambling},
journal={J. High Energy Phys.},
year={2015},
month={May},
day={26},
volume={2015},
number={5},
pages={132},
issn={1029-8479},
doi={10.1007/JHEP05(2015)132}
}

@article{Zhou19,
  title = {Operator dynamics in a Brownian quantum circuit},
  author = {Zhou, Tianci and Chen, Xiao},
  journal = {Phys. Rev. E},
  volume = {99},
  issue = {5},
  pages = {052212},
  numpages = {7},
  year = {2019},
  month = {May},
  publisher = {American Physical Society},
  doi = {10.1103/PhysRevE.99.052212}
}

@Article{Lashkari13,
author={Lashkari, Nima
and Stanford, Douglas
and Hastings, Matthew
and Osborne, Tobias
and Hayden, Patrick},
title={Towards the fast scrambling conjecture},
journal={J. High Energy Phys.},
year={2013},
month={Apr},
day={03},
volume={2013},
number={4},
pages={22},
issn={1029-8479},
doi={10.1007/JHEP04(2013)022}
}

@article{Fisher23,
author = {Fisher, Matthew P.A. and Khemani, Vedika and Nahum, Adam and Vijay, Sagar},
title = {Random Quantum Circuits},
journal = {Annu. Rev. Condens. Matter Phys.},
volume = {14},
number = {1},
pages = {335-379},
year = {2023}
}

@article{Seidelin06,
  title = {Microfabricated Surface-Electrode Ion Trap for Scalable Quantum Information Processing},
  author = {Seidelin, S. and Chiaverini, J. and Reichle, R. and Bollinger, J. J. and Leibfried, D. and Britton, J. and Wesenberg, J. H. and Blakestad, R. B. and Epstein, R. J. and Hume, D. B. and Itano, W. M. and Jost, J. D. and Langer, C. and Ozeri, R. and Shiga, N. and Wineland, D. J.},
  journal = {Phys. Rev. Lett.},
  volume = {96},
  issue = {25},
  pages = {253003},
  numpages = {4},
  year = {2006},
  month = {Jun},
  publisher = {American Physical Society},
  doi = {10.1103/PhysRevLett.96.253003}
}

@article{Bruzewicz19, title={Trapped-Ion Quantum Computing: {Progress} and Challenges}, volume={6}, DOI={10.1063/1.5088164}, number={2}, journal={Appl. Phys. Rev.}, author={Bruzewicz, Colin D. and Chiaverini, John and McConnell, Robert and Sage, Jeremy M.}, year={2019}, month={May}}

@article{Levine19,
  title = {Parallel Implementation of High-Fidelity Multiqubit Gates with Neutral Atoms},
  author = {Levine, Harry and Keesling, Alexander and Semeghini, Giulia and Omran, Ahmed and Wang, Tout T. and Ebadi, Sepehr and Bernien, Hannes and Greiner, Markus and Vuleti\ifmmode \acute{c}\else \'{c}\fi{}, Vladan and Pichler, Hannes and Lukin, Mikhail D.},
  journal = {Phys. Rev. Lett.},
  volume = {123},
  issue = {17},
  pages = {170503},
  numpages = {6},
  year = {2019},
  month = {Oct},
  publisher = {American Physical Society},
  doi = {10.1103/PhysRevLett.123.170503}
}

@Article{Zou20,
author={Zou, Jinhai
and Dong, Chuchu
and Wang, Hongjian
and Du, Tuanjie
and Luo, Zhengqian},
title={Towards visible-wavelength passively mode-locked lasers in all-fibre format},
journal={Light: Science {\&} Applications},
year={2020},
month={Apr},
day={14},
volume={9},
number={1},
pages={61},
issn={2047-7538},
doi={10.1038/s41377-020-0305-0}
}

@article{Shi22,
doi = {10.1088/2058-9565/ac18b8},
year = {2022},
month = {apr},
publisher = {IOP Publishing},
volume = {7},
number = {2},
pages = {023002},
author = {Xiao-Feng Shi},
title = {Quantum logic and entanglement by neutral {Rydberg} atoms: methods and fidelity},
journal = {Quantum Sci. Technol.}
}

@book{Fox06, place={Oxford}, title={Quantum Optics: An Introduction}, publisher={Oxford University Press}, author={Fox, Mark}, year={2006}}

@article{Saffman10,
  title = {Quantum information with {Rydberg} atoms},
  author = {Saffman, M. and Walker, T. G. and M\o{}lmer, K.},
  journal = {Rev. Mod. Phys.},
  volume = {82},
  issue = {3},
  pages = {2313--2363},
  numpages = {0},
  year = {2010},
  month = {Aug},
  publisher = {American Physical Society},
  doi = {10.1103/RevModPhys.82.2313}
}

@article{16_Saffman_Quantum,
doi = {10.1088/0953-4075/49/20/202001},
url = {https://dx.doi.org/10.1088/0953-4075/49/20/202001},
year = {2016},
month = {oct},
publisher = {IOP Publishing},
volume = {49},
number = {20},
pages = {202001},
author = {Saffman, M.},
title = {Quantum computing with atomic qubits and Rydberg interactions: progress and challenges},
journal = {J. Phys. B}
}

@article{19_Adams_Rydberg,
doi = {10.1088/1361-6455/ab52ef},
url = {https://dx.doi.org/10.1088/1361-6455/ab52ef},
year = {2019},
publisher = {IOP Publishing},
volume = {53},
number = {1},
pages = {012002},
author = {Adams, C.~S. and Pritchard, J.~D. and Shaffer, J.~P.},
title = {Rydberg atom quantum technologies},
journal = {J. Phys. B}}

@book{Silfvast04, place={Cambridge}, title={Laser Fundamentals}, publisher={Cambridge University Press}, author={Silfvast, William T.}, year={2004}}

@article{Innerhofer03,
author = {E. Innerhofer and T. S\"{u}dmeyer and F. Brunner and R. H\"{a}ring and A. Aschwanden and R. Paschotta and C. H\"{o}nninger and M. Kumkar and U. Keller},
journal = {Opt. Lett.},
number = {5},
pages = {367--369},
publisher = {Optica Publishing Group},
title = {{60-W average power in 810-fs pulses from a thin-disk Yb:YAG laser}},
volume = {28},
month = {Mar},
year = {2003},
doi = {10.1364/OL.28.000367},
}

@Article{Chew22,
author={Chew, Y.
and Tomita, T.
and Mahesh, T. P.
and Sugawa, S.
and de L{\'e}s{\'e}leuc, S.
and Ohmori, K.},
title={Ultrafast energy exchange between two single {Rydberg} atoms on a nanosecond timescale},
journal={Nat. Photon.},
year={2022},
month={Oct},
day={01},
volume={16},
number={10},
pages={724-729},
issn={1749-4893},
doi={10.1038/s41566-022-01047-2}
}

@Article{Xu22,
author={Xu, Peng
and Zhan, Ming-Sheng},
title={Ultrafast interaction between {Rydberg} atoms},
journal={Nat. Photon.},
year={2022},
month={Oct},
day={01},
volume={16},
number={10},
pages={673-674},
issn={1749-4893},
doi={10.1038/s41566-022-01074-z}
}

@Article{Abrams20,
author={Abrams, Deanna M.
and Didier, Nicolas
and Johnson, Blake R.
and Silva, Marcus P. da
and Ryan, Colm A.},
title={Implementation of {XY} entangling gates with a single calibrated pulse},
journal={Nat. Electron.},
year={2020},
month={Dec},
day={01},
volume={3},
number={12},
pages={744-750},
issn={2520-1131},
doi={10.1038/s41928-020-00498-1}
}

@article{Blais16,
  title = {Circuit quantum electrodynamics},
  author = {Blais, Alexandre and Grimsmo, Arne L. and Girvin, S. M. and Wallraff, Andreas},
  journal = {Rev. Mod. Phys.},
  volume = {93},
  issue = {2},
  pages = {025005},
  numpages = {72},
  year = {2021},
  month = {May},
  publisher = {American Physical Society},
  doi = {10.1103/RevModPhys.93.025005}
}

@book{Lambropoulos07, title={Fundamentals of Quantum Optics and Quantum Information}, publisher={Springer}, author={Lambropoulos, Peter and Petrosyan, David}, year={2007}}

@article{Tinkey22,
  title = {Transport-Enabled Entangling Gate for Trapped Ions},
  author = {Tinkey, Holly N. and Clark, Craig R. and Sawyer, Brian C. and Brown, Kenton R.},
  journal = {Phys. Rev. Lett.},
  volume = {128},
  issue = {5},
  pages = {050502},
  numpages = {6},
  year = {2022},
  month = {Jan},
  publisher = {American Physical Society},
  doi = {10.1103/PhysRevLett.128.050502}
}

@article{Monz11,
  title = {14-Qubit Entanglement: Creation and Coherence},
  author = {Monz, Thomas and Schindler, Philipp and Barreiro, Julio T. and Chwalla, Michael and Nigg, Daniel and Coish, William A. and Harlander, Maximilian and H\"ansel, Wolfgang and Hennrich, Markus and Blatt, Rainer},
  journal = {Phys. Rev. Lett.},
  volume = {106},
  issue = {13},
  pages = {130506},
  numpages = {4},
  year = {2011},
  month = {Mar},
  publisher = {American Physical Society},
  doi = {10.1103/PhysRevLett.106.130506}
}

@article{Kim10,
  title = {Surface-electrode point {Paul} trap},
  author = {Kim, Tony Hyun and Herskind, Peter F. and Kim, Taehyun and Kim, Jungsang and Chuang, Isaac L.},
  journal = {Phys. Rev. A},
  volume = {82},
  issue = {4},
  pages = {043412},
  numpages = {9},
  year = {2010},
  month = {Oct},
  publisher = {American Physical Society},
  doi = {10.1103/PhysRevA.82.043412}
}

@misc{Paschotta,
  author = {Paschotta, R.},
  publisher = {RP Photonics AG},
  title = {Frequency Doubling},
  year = {},
  month = {},
  howpublished = {RP Photonics Encyclopedia},
  note = {{Available online at \url{https://www.rp-photonics.com/frequency_doubling.html}}}
}

@article{Xia14,
title = {Nanosecond pulse generation in a graphene mode-locked erbium-doped fiber laser},
journal = {Opt. Commun.},
volume = {330},
pages = {147-150},
year = {2014},
issn = {0030-4018},
doi = {https://doi.org/10.1016/j.optcom.2014.05.048},
author = {Handing Xia and Heping Li and Zegao Wang and Yuanfu Chen and Xiaoxia Zhang and Xionggui Tang and Yong Liu}
}

@ARTICLE{Hong16,
  title    = "Guidelines for Designing Surface Ion Traps Using the Boundary
              Element Method",
  author   = "Hong, Seokjun and Lee, Minjae and Cheon, Hongjin and Kim, Taehyun
              and Cho, Dong-Il Dan",
  journal  = "Sensors (Basel)",
  volume   =  16,
  number   =  5,
  month    =  apr,
  year     =  2016,
  address  = "Switzerland",
  language = "en"
}

@article{Dietrich10,
  title = {Hyperfine and optical barium ion qubits},
  author = {Dietrich, M. R. and Kurz, N. and Noel, T. and Shu, G. and Blinov, B. B.},
  journal = {Phys. Rev. A},
  volume = {81},
  issue = {5},
  pages = {052328},
  numpages = {4},
  year = {2010},
  month = {May},
  publisher = {American Physical Society},
  doi = {10.1103/PhysRevA.81.052328}
}

@article{Hughes20,
  title = {Benchmarking a High-Fidelity Mixed-Species Entangling Gate},
  author = {Hughes, A. C. and Sch\"afer, V. M. and Thirumalai, K. and Nadlinger, D. P. and Woodrow, S. R. and Lucas, D. M. and Ballance, C. J.},
  journal = {Phys. Rev. Lett.},
  volume = {125},
  issue = {8},
  pages = {080504},
  numpages = {7},
  year = {2020},
  month = {Aug},
  publisher = {American Physical Society},
  doi = {10.1103/PhysRevLett.125.080504}
}

@Article{Wright19,
author={Wright, K.
and Beck, K. M.
and Debnath, S.
and Amini, J. M.
and Nam, Y.
and Grzesiak, N.
and Chen, J.-S.
and Pisenti, N. C.
and Chmielewski, M.
and Collins, C.
and Hudek, K. M.
and Mizrahi, J.
and Wong-Campos, J. D.
and Allen, S.
and Apisdorf, J.
and Solomon, P.
and Williams, M.
and Ducore, A. M.
and Blinov, A.
and Kreikemeier, S. M.
and Chaplin, V.
and Keesan, M.
and Monroe, C.
and Kim, J.},
title={Benchmarking an 11-qubit quantum computer},
journal={Nat. Commun.},
year={2019},
month={Nov},
day={29},
volume={10},
number={1},
pages={5464},
issn={2041-1723},
doi={10.1038/s41467-019-13534-2}
}

@Article{Gaetan09,
author={Ga{\"e}tan, Alpha
and Miroshnychenko, Yevhen
and Wilk, Tatjana
and Chotia, Amodsen
and Viteau, Matthieu
and Comparat, Daniel
and Pillet, Pierre
and Browaeys, Antoine
and Grangier, Philippe},
title={Observation of collective excitation of two individual atoms in the {Rydberg} blockade regime},
journal={Nat. Phys.},
year={2009},
month={Feb},
day={01},
volume={5},
number={2},
pages={115-118},
issn={1745-2481},
doi={10.1038/nphys1183}
}

@article{Bartels08,
author = {A. Bartels and D. Heinecke and S. A. Diddams},
journal = {Opt. Lett.},
number = {16},
pages = {1905--1907},
publisher = {Optica Publishing Group},
title = {Passively mode-locked 10 {GHz} femtosecond {Ti:sapphire} laser},
volume = {33},
month = {Aug},
year = {2008},
doi = {10.1364/OL.33.001905},
}

@misc{PaschottaSync,
  author = {Paschotta, R.},
  publisher = {RP Photonics AG},
  title = {Synchronization of Lasers},
  year = {},
  month = {},
  howpublished = {RP Photonics Encyclopedia},
  note = {{Available online at \url{https://www.rp-photonics.com/synchronization_of_lasers.html}}
}}

@article{Besse18,
  title = {Single-Shot Quantum Nondemolition Detection of Individual Itinerant Microwave Photons},
  author = {Besse, Jean-Claude and Gasparinetti, Simone and Collodo, Michele C. and Walter, Theo and Kurpiers, Philipp and Pechal, Marek and Eichler, Christopher and Wallraff, Andreas},
  journal = {Phys. Rev. X},
  volume = {8},
  issue = {2},
  pages = {021003},
  numpages = {9},
  year = {2018},
  month = {Apr},
  publisher = {American Physical Society},
  doi = {10.1103/PhysRevX.8.021003}
}

@Article{Ebadi21,
author={Ebadi, Sepehr
and Wang, Tout T.
and Levine, Harry
and Keesling, Alexander
and Semeghini, Giulia
and Omran, Ahmed
and Bluvstein, Dolev
and Samajdar, Rhine
and Pichler, Hannes
and Ho, Wen Wei
and Choi, Soonwon
and Sachdev, Subir
and Greiner, Markus
and Vuleti{\'{c}}, Vladan
and Lukin, Mikhail D.},
title={Quantum phases of matter on a 256-atom programmable quantum simulator},
journal={Nature},
year={2021},
month={Jul},
day={01},
volume={595},
number={7866},
pages={227-232},
issn={1476-4687},
doi={10.1038/s41586-021-03582-4}
}

@article{Linden10,
  title = {How Small Can Thermal Machines Be? The Smallest Possible Refrigerator},
  author = {Linden, Noah and Popescu, Sandu and Skrzypczyk, Paul},
  journal = {Phys. Rev. Lett.},
  volume = {105},
  issue = {13},
  pages = {130401},
  numpages = {4},
  year = {2010},
  month = {Sep},
  publisher = {American Physical Society},
  doi = {10.1103/PhysRevLett.105.130401}}

@article{DiVincenzo00,
  title = {The Physical Implementation of Quantum Computation},
  author = {DiVincenzo, D. P.},
  year = {2000},
  journal = {Fortschr. Phys.},
  volume = {48},
  number = {9-11},
  pages = {771--783},
  doi = {10.1002/1521-3978(200009)48:9/11<771::AID-PROP771>3.0.CO;2-E}
}

@Article{Novikov16,
author={Novikov, S.
and Sweeney, T.
and Robinson, J. E.
and Premaratne, S. P.
and Suri, B.
and Wellstood, F. C.
and Palmer, B. S.},
title={Raman coherence in a circuit quantum electrodynamics lambda system},
journal={Nat. Phys.},
year={2016},
month={Jan},
day={01},
volume={12},
number={1},
pages={75-79},
issn={1745-2481},
doi={10.1038/nphys3537}
}

@article{Krantz19, title={A quantum engineer’s guide to superconducting qubits}, volume={6}, DOI={10.1063/1.5089550}, number={2}, journal={Appl. Phys. Rev.}, author={Krantz, P. and Kjaergaard, M. and Yan, F. and Orlando, T. P. and Gustavsson, S. and Oliver, W. D.}, year={2019}, month={Jun}}

@article{Yan18,
  title = {Tunable Coupling Scheme for Implementing High-Fidelity Two-Qubit Gates},
  author = {Yan, Fei and Krantz, Philip and Sung, Youngkyu and Kjaergaard, Morten and Campbell, Daniel L. and Orlando, Terry P. and Gustavsson, Simon and Oliver, William D.},
  journal = {Phys. Rev. Appl.},
  volume = {10},
  issue = {5},
  pages = {054062},
  numpages = {9},
  year = {2018},
  month = {Nov},
  publisher = {American Physical Society},
  doi = {10.1103/PhysRevApplied.10.054062}
}

@article{
Ye24,
author = {Yufeng Ye  and Jeremy B. Kline  and Sean Chen  and Alec Yen  and Kevin P. O’Brien },
title = {Ultrafast superconducting qubit readout with the quarton coupler},
journal = {Sci. Adv.},
volume = {10},
number = {41},
pages = {eado9094},
year = {2024},
doi = {10.1126/sciadv.ado9094}}

@misc{PaschottaPG,
  author = {Paschotta, R.},
  publisher = {RP Photonics AG},
  title = {Pulse Generation},
  year = {},
  month = {},
  howpublished = {RP Photonics Encyclopedia},
  note = {{Available online at \url{https://www.rp-photonics.com/pulse_generation.html}
}}}

@article{YungerHalpern20,
  title = {Fundamental limitations on photoisomerization from thermodynamic resource theories},
  author = {Yunger Halpern, Nicole and Limmer, David T.},
  journal = {Phys. Rev. A},
  volume = {101},
  issue = {4},
  pages = {042116},
  numpages = {15},
  year = {2020},
  month = {Apr},
  publisher = {American Physical Society},
  doi = {10.1103/PhysRevA.101.042116}
}

@Article{Verstraete09,
author={Verstraete, Frank
and Wolf, Michael M.
and Ignacio Cirac, J.},
title={Quantum computation and quantum-state engineering driven by dissipation},
journal={Nat. Phys.},
year={2009},
month={Sep},
day={01},
volume={5},
number={9},
pages={633-636},
issn={1745-2481},
doi={10.1038/nphys1342}
}

@article{Ghasemian23,
author = {Ebrahim Ghasemian},
journal = {J. Opt. Soc. Am. B},
number = {2},
pages = {247--259},
publisher = {Optica Publishing Group},
title = {Dissipative quantum computation and quantum state preparation based on {BEC} qubits},
volume = {40},
month = {Feb},
year = {2023},
doi = {10.1364/JOSAB.470605}
}

@article{Zapusek23,
doi = {10.1088/2058-9565/ac98dd},
year = {2022},
month = {nov},
publisher = {IOP Publishing},
volume = {8},
number = {1},
pages = {015001},
author = {Zapusek, Elias and Javadi, Alisa and Reiter, Florentin},
title = {Nonunitary gate operations by dissipation engineering},
journal = {Quantum Sci. Technol.}
}

@article{GelbwaserKlimovsky14,
  title = {Heat-machine control by quantum-state preparation: From quantum engines to refrigerators},
  author = {Gelbwaser-Klimovsky, D. and Kurizki, G.},
  journal = {Phys. Rev. E},
  volume = {90},
  issue = {2},
  pages = {022102},
  numpages = {12},
  year = {2014},
  month = {Aug},
  publisher = {American Physical Society},
  doi = {10.1103/PhysRevE.90.022102}
}

@article{Roulet17,
  title = {Autonomous rotor heat engine},
  author = {Roulet, Alexandre and Nimmrichter, Stefan and Arrazola, Juan Miguel and Seah, Stella and Scarani, Valerio},
  journal = {Phys. Rev. E},
  volume = {95},
  issue = {6},
  pages = {062131},
  numpages = {13},
  year = {2017},
  month = {Jun},
  publisher = {American Physical Society},
  doi = {10.1103/PhysRevE.95.062131}
}

@Article{Correa14,
author={Correa, Luis A.
and Palao, Jos{\'e} P.
and Alonso, Daniel
and Adesso, Gerardo},
title={Quantum-enhanced absorption refrigerators},
journal={Sci. Rep.},
year={2014},
month={Feb},
day={04},
volume={4},
number={1},
pages={3949},
issn={2045-2322},
doi={10.1038/srep03949}
}

@article{Koski15,
  title = {On-Chip {Maxwell's} Demon as an Information-Powered Refrigerator},
  author = {Koski, J. V. and Kutvonen, A. and Khaymovich, I. M. and Ala-Nissila, T. and Pekola, J. P.},
  journal = {Phys. Rev. Lett.},
  volume = {115},
  issue = {26},
  pages = {260602},
  numpages = {5},
  year = {2015},
  month = {Dec},
  publisher = {American Physical Society},
  doi = {10.1103/PhysRevLett.115.260602}
}

@article{Strasberg18,
  title = {Fermionic reaction coordinates and their application to an autonomous {Maxwell} demon in the strong-coupling regime},
  author = {Strasberg, Philipp and Schaller, Gernot and Schmidt, Thomas L. and Esposito, Massimiliano},
  journal = {Phys. Rev. B},
  volume = {97},
  issue = {20},
  pages = {205405},
  numpages = {16},
  year = {2018},
  month = {May},
  publisher = {American Physical Society},
  doi = {10.1103/PhysRevB.97.205405}
}

@article{Brunner12,
  title = {Virtual qubits, virtual temperatures, and the foundations of thermodynamics},
  author = {Brunner, Nicolas and Linden, Noah and Popescu, Sandu and Skrzypczyk, Paul},
  journal = {Phys. Rev. E},
  volume = {85},
  issue = {5},
  pages = {051117},
  numpages = {14},
  year = {2012},
  month = {May},
  publisher = {American Physical Society},
  doi = {10.1103/PhysRevE.85.051117}
}

@article{Molmer99,
  title = {Multiparticle Entanglement of Hot Trapped Ions},
  author = {M\o{}lmer, Klaus and S\o{}rensen, Anders},
  journal = {Phys. Rev. Lett.},
  volume = {82},
  issue = {9},
  pages = {1835--1838},
  numpages = {0},
  year = {1999},
  month = {Mar},
  publisher = {American Physical Society},
  doi = {10.1103/PhysRevLett.82.1835}
}

@article{Suleimen24,
  title = {Surface trap with adjustable ion couplings for scalable and parallel gates},
  author = {Suleimen, Y. and Podlesnyy, A. and Akopyan, L. A. and Sterligov, N. and Lakhmanskaya, O. and Anikin, E. and Matveev, A. and Lakhmanskiy, K.},
  journal = {Phys. Rev. A},
  volume = {109},
  issue = {2},
  pages = {022605},
  numpages = {17},
  year = {2024},
  month = {Feb},
  publisher = {American Physical Society},
  doi = {10.1103/PhysRevA.109.022605}
}

@article{Schindler13,
doi = {10.1088/1367-2630/15/12/123012},
year = {2013},
month = {dec},
publisher = {IOP Publishing},
volume = {15},
number = {12},
pages = {123012},
author = {Schindler, Philipp and Nigg, Daniel and Monz, Thomas and Barreiro, Julio T and Martinez, Esteban and Wang, Shannon X and Quint, Stephan and Brandl, Matthias F and Nebendahl, Volckmar and Roos, Christian F and Chwalla, Michael and Hennrich, Markus and Blatt, Rainer},
title = {A quantum information processor with trapped ions},
journal = {New J. Phys.}
}

@Article{Ospelkaus11,
author={Ospelkaus, C.
and Warring, U.
and Colombe, Y.
and Brown, K. R.
and Amini, J. M.
and Leibfried, D.
and Wineland, D. J.},
title={Microwave quantum logic gates for trapped ions},
journal={Nature},
year={2011},
month={Aug},
day={01},
volume={476},
number={7359},
pages={181-184},
issn={1476-4687},
doi={10.1038/nature10290}
}

@article{Campbell10,
  title = {Ultrafast Gates for Single Atomic Qubits},
  author = {Campbell, W. C. and Mizrahi, J. and Quraishi, Q. and Senko, C. and Hayes, D. and Hucul, D. and Matsukevich, D. N. and Maunz, P. and Monroe, C.},
  journal = {Phys. Rev. Lett.},
  volume = {105},
  issue = {9},
  pages = {090502},
  numpages = {4},
  year = {2010},
  month = {Aug},
  publisher = {American Physical Society},
  doi = {10.1103/PhysRevLett.105.090502}
}

@article{Ballance16,
  title = {High-Fidelity Quantum Logic Gates Using Trapped-Ion Hyperfine Qubits},
  author = {Ballance, C. J. and Harty, T. P. and Linke, N. M. and Sepiol, M. A. and Lucas, D. M.},
  journal = {Phys. Rev. Lett.},
  volume = {117},
  issue = {6},
  pages = {060504},
  numpages = {6},
  year = {2016},
  month = {Aug},
  publisher = {American Physical Society},
  doi = {10.1103/PhysRevLett.117.060504}
}

@Article{Benhelm08,
author={Benhelm, Jan
and Kirchmair, Gerhard
and Roos, Christian F.
and Blatt, Rainer},
title={Towards fault-tolerant quantum computing with trapped ions},
journal={Nat. Phys.},
year={2008},
month={Jun},
day={01},
volume={4},
number={6},
pages={463-466},
issn={1745-2481},
doi={10.1038/nphys961}
}

@article{Saskin19,
  title = {Narrow-Line Cooling and Imaging of Ytterbium Atoms in an Optical Tweezer Array},
  author = {Saskin, S. and Wilson, J. T. and Grinkemeyer, B. and Thompson, J. D.},
  journal = {Phys. Rev. Lett.},
  volume = {122},
  issue = {14},
  pages = {143002},
  numpages = {6},
  year = {2019},
  month = {Apr},
  publisher = {American Physical Society},
  doi = {10.1103/PhysRevLett.122.143002}
}

@article{Liu15,
title = {Generation of microseconds-duration square pulses in a passively mode-locked fiber laser},
journal = {Opt. Commun.},
volume = {356},
pages = {416-420},
year = {2015},
issn = {0030-4018},
doi = {https://doi.org/10.1016/j.optcom.2015.08.032},
author = {Tonghui Liu and Dongfang Jia and Ying Liu and Zhaoying Wang and Tianxin Yang}
}

@article{Prech24,
  title = {Optimal Time Estimation and the Clock Uncertainty Relation for Stochastic Processes},
  author = {Prech, Kacper and Landi, Gabriel T. and Meier, Florian and Nurgalieva, Nuriya and Potts, Patrick P. and Silva, Ralph and Mitchison, Mark T.},
  journal = {Phys. Rev. X},
  volume = {15},
  issue = {3},
  pages = {031068},
  numpages = {37},
  year = {2025},
  month = {Sep},
  publisher = {American Physical Society},
  doi = {10.1103/rpls-mp8z},
  url = {https://link.aps.org/doi/10.1103/rpls-mp8z}
}

@article{Vinjanampathy16,
author = {Sai Vinjanampathy and Janet Anders},
title = {Quantum thermodynamics},
journal = {Contemp. Phys.},
volume = {57},
number = {4},
pages = {545--579},
year = {2016},
publisher = {Taylor \& Francis}
}

@BOOK{Grumbling19,
  editor    = "Emily Grumbling and Mark Horowitz",
  title     = "Quantum Computing: Progress and Prospects",
  isbn      = "978-0-309-47969-1",
  doi       = "10.17226/25196",
  year      = 2019,
  publisher = "The National Academies Press",
  address   = "Washington, DC"
}

@Article{Krinner19,
author={Krinner, S.
and Storz, S.
and Kurpiers, P.
and Magnard, P.
and Heinsoo, J.
and Keller, R.
and L{\"u}tolf, J.
and Eichler, C.
and Wallraff, A.},
title={Engineering cryogenic setups for 100-qubit scale superconducting circuit systems},
journal={Eur. Phys. J. Quantum Technol.},
year={2019},
month={May},
day={28},
volume={6},
number={1},
pages={2},
issn={2196-0763}
}

@article{Elouard23,
  title = {Extending the Laws of Thermodynamics for Arbitrary Autonomous Quantum Systems},
  author = {Elouard, Cyril and Lombard Latune, Camille},
  journal = {PRX Quantum},
  volume = {4},
  issue = {2},
  pages = {020309},
  numpages = {19},
  year = {2023},
  month = {Apr},
  publisher = {American Physical Society}
}

@article{23_Xuereb_Impact,
  title = {Impact of imperfect timekeeping on quantum control},
  author = {Xuereb, Jake and Erker, Paul and Meier, Florian and Mitchison, Mark T. and Huber, Marcus},
  journal = {Phys. Rev. Lett.},
  volume = {131},
  issue = {16},
  pages = {160204},
  numpages = {7},
  year = {2023},
  month = {Oct},
  publisher = {American Physical Society},
  doi = {10.1103/PhysRevLett.131.160204},
  url = {https://link.aps.org/doi/10.1103/PhysRevLett.131.160204}
}

@article{23_Meier_Fundamental,
  title = {Fundamental Accuracy-Resolution Trade-Off for Timekeeping Devices},
  author = {Meier, Florian and Schwarzhans, Emanuel and Erker, Paul and Huber, Marcus},
  journal = {Phys. Rev. Lett.},
  volume = {131},
  issue = {22},
  pages = {220201},
  numpages = {6},
  year = {2023},
  month = {Nov},
  publisher = {American Physical Society},
  doi = {10.1103/PhysRevLett.131.220201},
  url = {https://link.aps.org/doi/10.1103/PhysRevLett.131.220201}
}

@article{23_Fisher_Random,
   author = "Fisher, Matthew P.~A. and Khemani, Vedika and Nahum, Adam and Vijay, Sagar",
   title = "Random quantum circuits", 
   journal= "Annu. Rev. Condens. Matter Phys.",
   year = "2023",
   volume = "14",
   number = "Volume 14, 2023",
   pages = "335-379",
   doi = "https://doi.org/10.1146/annurev-conmatphys-031720-030658",
   url = "https://www.annualreviews.org/content/journals/10.1146/annurev-conmatphys-031720-030658",
   publisher = "Annual Reviews",
   issn = "1947-5462"
   }

@ARTICLE{21_Lebreuilly_Autonomous,
       author = {{Lebreuilly}, Jos{\'e} and {Noh}, Kyungjoo and {Wang}, Chiao-Hsuan and {Girvin}, Steven M. and {Jiang}, Liang},
        title = "{Autonomous quantum error correction and quantum computation}",
      journal = {arXiv e-prints},
         year = 2021,
        month = mar,
          eid = {arXiv:2103.05007},
        pages = {arXiv:2103.05007},
          doi = {10.48550/arXiv.2103.05007},
archivePrefix = {arXiv},
       eprint = {2103.05007},
 primaryClass = {quant-ph},
       adsurl = {https://ui.adsabs.harvard.edu/abs/2021arXiv210305007L}
}

@book{NielsenC10,
	Author = {Nielsen, Michael A. and Chuang, Isaac L.},
	Publisher = {Cambridge University Press},
	Title = {{Quantum Computation and Quantum Information}},
	Year = {2010}}

@article{Mundada19,
  title = {Suppression of Qubit Crosstalk in a Tunable Coupling Superconducting Circuit},
  author = {Mundada, Pranav and Zhang, Gengyan and Hazard, Thomas and Houck, Andrew},
  journal = {Phys. Rev. Appl.},
  volume = {12},
  issue = {5},
  pages = {054023},
  numpages = {10},
  year = {2019},
  month = {Nov},
  publisher = {American Physical Society}
}

@article{Wu21,
year = {2021},
month = {feb},
publisher = {Chinese Physical Society and IOP Publishing Ltd},
volume = {30},
number = {2},
pages = {020305},
author = {Wu, Xiaoling and Liang, Xinhui and Tian, Yaoqi and Yang, Fan and Chen, Cheng and Liu, Yong-Chun and Tey, Meng Khoon and You, Li},
title = {A concise review of Rydberg atom based quantum computation and quantum simulation*},
journal = {Chin. Phys. B}
}

@article{15_Ludlow_Optical,
  title = {Optical atomic clocks},
  author = {Ludlow, Andrew D. and Boyd, Martin M. and Ye, Jun and Peik, E. and Schmidt, P. O.},
  journal = {Rev. Mod. Phys.},
  volume = {87},
  issue = {2},
  pages = {637--701},
  numpages = {65},
  year = {2015},
  month = {Jun},
  publisher = {American Physical Society},
  doi = {10.1103/RevModPhys.87.637},
  url = {https://link.aps.org/doi/10.1103/RevModPhys.87.637}
}

@Article{Meier25,
author={Meier, Florian
and Minoguchi, Yuri
and Sundelin, Simon
and Apollaro, Tony J. G.
and Erker, Paul
and Gasparinetti, Simone
and Huber, Marcus},
title={Precision is not limited by the second law of thermodynamics},
journal={Nat. Phys.},
year={2025},
month={Jul},
day={01},
volume={21},
number={7},
pages={1147-1152},
issn={1745-2481},
doi={10.1038/s41567-025-02929-2},
url={https://doi.org/10.1038/s41567-025-02929-2}
}

@misc{PaschottaLamp,
  author = {Paschotta, R.},
  publisher = {RP Photonics AG},
  title = {Lamp-pumped Lasers},
  year = {2005},
  month = {},
  howpublished = {RP Photonics Encyclopedia},
  note = {{Available online at \url{https://www.rp-photonics.com/lamp_pumped_lasers.html}}}}

@Article{Kono18,
author={Kono, S.
and Koshino, K.
and Tabuchi, Y.
and Noguchi, A.
and Nakamura, Y.},
title={Quantum non-demolition detection of an itinerant microwave photon},
journal={Nat. Phys.},
year={2018},
month={Jun},
day={01},
volume={14},
number={6},
pages={546-549},
issn={1745-2481},
doi={10.1038/s41567-018-0066-3},
url={https://doi.org/10.1038/s41567-018-0066-3}
}

@Article{Hacker16,
author={Hacker, Bastian
and Welte, Stephan
and Rempe, Gerhard
and Ritter, Stephan},
title={A photon--photon quantum gate based on a single atom in an optical resonator},
journal={Nature},
year={2016},
month={Aug},
day={01},
volume={536},
number={7615},
pages={193-196},
issn={1476-4687},
doi={10.1038/nature18592},
url={https://doi.org/10.1038/nature18592}
}

@article{Duan04,
  title = {Scalable Photonic Quantum Computation through Cavity-Assisted Interactions},
  author = {Duan, L.-M. and Kimble, H. J.},
  journal = {Phys. Rev. Lett.},
  volume = {92},
  issue = {12},
  pages = {127902},
  numpages = {4},
  year = {2004},
  month = {Mar},
  publisher = {American Physical Society},
  doi = {10.1103/PhysRevLett.92.127902},
  url = {https://link.aps.org/doi/10.1103/PhysRevLett.92.127902}
}

@article{McDonald23,
  title = {Third quantization of open quantum systems: Dissipative symmetries and connections to phase-space and Keldysh field-theory formulations},
  author = {McDonald, Alexander and Clerk, Aashish A.},
  journal = {Phys. Rev. Res.},
  volume = {5},
  issue = {3},
  pages = {033107},
  numpages = {20},
  year = {2023},
  month = {Aug},
  publisher = {American Physical Society},
  doi = {10.1103/PhysRevResearch.5.033107},
  url = {https://link.aps.org/doi/10.1103/PhysRevResearch.5.033107}
}

@article{Gardiner85,
  title = {Input and output in damped quantum systems: Quantum stochastic differential equations and the master equation},
  author = {Gardiner, C. W. and Collett, M. J.},
  journal = {Phys. Rev. A},
  volume = {31},
  issue = {6},
  pages = {3761--3774},
  numpages = {0},
  year = {1985},
  month = {Jun},
  publisher = {American Physical Society},
  doi = {10.1103/PhysRevA.31.3761},
  url = {https://link.aps.org/doi/10.1103/PhysRevA.31.3761}
}

@Article{Inomata16,
author={Inomata, Kunihiro
and Lin, Zhirong
and Koshino, Kazuki
and Oliver, William D.
and Tsai, Jaw-Shen
and Yamamoto, Tsuyoshi
and Nakamura, Yasunobu},
title={Single microwave-photon detector using an artificial $\Lambda$-type three-level system},
journal={Nat. Commun.},
year={2016},
month={Jul},
day={25},
volume={7},
number={1},
pages={12303},
issn={2041-1723},
doi={10.1038/ncomms12303},
url={https://doi.org/10.1038/ncomms12303}
}

@article{Schwarzhans26,
  title = {Quantum Detectors as Autonomous Machines: Assessing the Nonequilibrium Thermodynamics of Information Acquisition},
  author = {Schwarzhans, Emanuel and Apollaro, Tony J. G. and Khomchenko, Ilia and Lock, Maximilian P. E. and Mitchison, Mark T. and Huber, Marcus},
  journal = {PRX Quantum},
  volume = {7},
  issue = {3},
  pages = {033001},
  numpages = {14},
  year = {2026},
  month = {Jul},
  publisher = {American Physical Society},
  doi = {10.1103/wm5p-tjtg},
  url = {https://link.aps.org/doi/10.1103/wm5p-tjtg}
}

@article{deClercq16,
  title = {Parallel Transport Quantum Logic Gates with Trapped Ions},
  author = {de Clercq, Ludwig E. and Lo, Hsiang-Yu and Marinelli, Matteo and Nadlinger, David and Oswald, Robin and Negnevitsky, Vlad and Kienzler, Daniel and Keitch, Ben and Home, Jonathan P.},
  journal = {Phys. Rev. Lett.},
  volume = {116},
  issue = {8},
  pages = {080502},
  numpages = {5},
  year = {2016},
  month = {Feb},
  publisher = {American Physical Society},
  doi = {10.1103/PhysRevLett.116.080502},
  url = {https://link.aps.org/doi/10.1103/PhysRevLett.116.080502}
}

@article{Leibfreid07,
  title = {Transport quantum logic gates for trapped ions},
  author = {Leibfried, D. and Knill, E. and Ospelkaus, C. and Wineland, D. J.},
  journal = {Phys. Rev. A},
  volume = {76},
  issue = {3},
  pages = {032324},
  numpages = {12},
  year = {2007},
  month = {Sep},
  publisher = {American Physical Society},
  doi = {10.1103/PhysRevA.76.032324},
  url = {https://link.aps.org/doi/10.1103/PhysRevA.76.032324}
}

@ARTICLE{Sterk24,
       author = {{Sterk}, J.~D. and {Blain}, M.~G. and {Delaney}, M. and {Haltli}, R. and {Heller}, E. and {Holterhoff}, A.~L. and {Jennings}, T. and {Jimenez}, N. and {Kozhanov}, A. and {Meinelt}, Z. and {Ou}, E. and {Van Der Wall}, J. and {Noel}, C. and {Stick}, D.},
        title = "{Multi-junction surface ion trap for quantum computing}",
      journal = {arXiv e-prints},
         year = 2024,
        month = feb,
          eid = {arXiv:2403.00208},
        pages = {arXiv:2403.00208},
          doi = {10.48550/arXiv.2403.00208},
archivePrefix = {arXiv},
       eprint = {2403.00208},
 primaryClass = {quant-ph},
       adsurl = {https://ui.adsabs.harvard.edu/abs/2024arXiv240300208S}
}

@article{Wilk10,
  title = {Entanglement of Two Individual Neutral Atoms Using Rydberg Blockade},
  author = {Wilk, T. and Ga\"etan, A. and Evellin, C. and Wolters, J. and Miroshnychenko, Y. and Grangier, P. and Browaeys, A.},
  journal = {Phys. Rev. Lett.},
  volume = {104},
  issue = {1},
  pages = {010502},
  numpages = {4},
  year = {2010},
  month = {Jan},
  publisher = {American Physical Society},
  doi = {10.1103/PhysRevLett.104.010502},
  url = {https://link.aps.org/doi/10.1103/PhysRevLett.104.010502}
}

@article{Odelin19,
  title = {Shortcuts to adiabaticity: Concepts, methods, and applications},
  author = {Gu\'ery-Odelin, D. and Ruschhaupt, A. and Kiely, A. and Torrontegui, E. and Mart\'{\i}nez-Garaot, S. and Muga, J. G.},
  journal = {Rev. Mod. Phys.},
  volume = {91},
  issue = {4},
  pages = {045001},
  numpages = {54},
  year = {2019},
  month = {Oct},
  publisher = {American Physical Society},
  doi = {10.1103/RevModPhys.91.045001},
  url = {https://link.aps.org/doi/10.1103/RevModPhys.91.045001}
}

@article{Page83,
  title = {Evolution without evolution: Dynamics described by stationary observables},
  author = {Page, Don N. and Wootters, William K.},
  journal = {Phys. Rev. D},
  volume = {27},
  issue = {12},
  pages = {2885--2892},
  numpages = {0},
  year = {1983},
  month = {Jun},
  publisher = {American Physical Society},
  doi = {10.1103/PhysRevD.27.2885},
  url = {https://link.aps.org/doi/10.1103/PhysRevD.27.2885}
}

@article{Hahn50,
  title = {Spin Echoes},
  author = {Hahn, E. L.},
  journal = {Phys. Rev.},
  volume = {80},
  issue = {4},
  pages = {580--594},
  numpages = {0},
  year = {1950},
  month = {Nov},
  publisher = {American Physical Society},
  doi = {10.1103/PhysRev.80.580},
  url = {https://link.aps.org/doi/10.1103/PhysRev.80.580}
}

@article{Leghtas13,
  title = {Hardware-Efficient Autonomous Quantum Memory Protection},
  author = {Leghtas, Zaki and Kirchmair, Gerhard and Vlastakis, Brian and Schoelkopf, Robert J. and Devoret, Michel H. and Mirrahimi, Mazyar},
  journal = {Phys. Rev. Lett.},
  volume = {111},
  issue = {12},
  pages = {120501},
  numpages = {5},
  year = {2013},
  month = {Sep},
  publisher = {American Physical Society},
  doi = {10.1103/PhysRevLett.111.120501},
  url = {https://link.aps.org/doi/10.1103/PhysRevLett.111.120501}
}

@article{Gertler21,
   title={Protecting a bosonic qubit with autonomous quantum error correction},
   volume={590},
   ISSN={1476-4687},
   url={http://dx.doi.org/10.1038/s41586-021-03257-0},
   DOI={10.1038/s41586-021-03257-0},
   number={7845},
   journal={Nature},
   publisher={Springer Science and Business Media LLC},
   author={Gertler, Jeffrey M. and Baker, Brian and Li, Juliang and Shirol, Shruti and Koch, Jens and Wang, Chen},
   year={2021},
   month=feb, pages={243–248} }

@article{Perez20,
   title={Improved autonomous error correction using variable dissipation in small logical qubit architectures},
   volume={6},
   ISSN={2058-9565},
   url={http://dx.doi.org/10.1088/2058-9565/abc3cb},
   DOI={10.1088/2058-9565/abc3cb},
   number={1},
   journal={Quantum Sci. Technol.},
   publisher={IOP Publishing},
   author={Pérez, David Rodríguez and Kapit, Eliot},
   year={2020},
   month=nov, pages={015006} }

@article{Mahesh25,
author = {T. P. Mahesh and Takuya Matsubara and Yuki Torii Chew and Takafumi Tomita and Sylvain de L\'{e}s\'{e}leuc and Kenji Ohmori},
journal = {Opt. Lett.},
number = {2},
pages = {403--406},
publisher = {Optica Publishing Group},
title = {Generation of 480 nm picosecond pulses for ultrafast excitation of Rydberg atoms},
volume = {50},
month = {Jan},
year = {2025},
url = {https://opg.optica.org/ol/abstract.cfm?URI=ol-50-2-403},
doi = {10.1364/OL.538707},
}

@misc{rpmc_nps_laser,
  author       = {{RPMC Lasers}},
  title        = {{NPS: Narrowband Picosecond Mode-Locked Laser}},
  howpublished = {\url{https://www.rpmclasers.com/product/nps-narrowband-picosecond-mode-locked-laser/}},
  year         = {2026},
  note         = {Accessed: 2026-04-14}
}

@misc{PaschottaQsw,
  author = {Paschotta, R.},
  publisher = {RP Photonics AG},
  title = {Q-switching},
  year = {2021},
  howpublished = {RP Photonics Encyclopedia},
  note = {Available online at \url{https://www.rp-photonics.com/q_switching.html}},
  url = {https://www.rp-photonics.com/q_switching.html},
  doi = {10.61835/fdh},
  urldate = {2026-04-16}
}

@Article{Arute19,
author={Arute, Frank
and Arya, Kunal
and Babbush, Ryan
and Bacon, Dave
and Bardin, Joseph C.
and Barends, Rami
and Biswas, Rupak
and Boixo, Sergio
and Brandao, Fernando G. S. L.
and Buell, David A.
and Burkett, Brian
and Chen, Yu
and Chen, Zijun
and Chiaro, Ben
and Collins, Roberto
and Courtney, William
and Dunsworth, Andrew
and Farhi, Edward
and Foxen, Brooks
and Fowler, Austin
and Gidney, Craig
and Giustina, Marissa
and Graff, Rob
and Guerin, Keith
and Habegger, Steve
and Harrigan, Matthew P.
and Hartmann, Michael J.
and Ho, Alan
and Hoffmann, Markus
and Huang, Trent
and Humble, Travis S.
and Isakov, Sergei V.
and Jeffrey, Evan
and Jiang, Zhang
and Kafri, Dvir
and Kechedzhi, Kostyantyn
and Kelly, Julian
and Klimov, Paul V.
and Knysh, Sergey
and Korotkov, Alexander
and Kostritsa, Fedor
and Landhuis, David
and Lindmark, Mike
and Lucero, Erik
and Lyakh, Dmitry
and Mandr{\`a}, Salvatore
and McClean, Jarrod R.
and McEwen, Matthew
and Megrant, Anthony
and Mi, Xiao
and Michielsen, Kristel
and Mohseni, Masoud
and Mutus, Josh
and Naaman, Ofer
and Neeley, Matthew
and Neill, Charles
and Niu, Murphy Yuezhen
and Ostby, Eric
and Petukhov, Andre
and Platt, John C.
and Quintana, Chris
and Rieffel, Eleanor G.
and Roushan, Pedram
and Rubin, Nicholas C.
and Sank, Daniel
and Satzinger, Kevin J.
and Smelyanskiy, Vadim
and Sung, Kevin J.
and Trevithick, Matthew D.
and Vainsencher, Amit
and Villalonga, Benjamin
and White, Theodore
and Yao, Z. Jamie
and Yeh, Ping
and Zalcman, Adam
and Neven, Hartmut
and Martinis, John M.},
title={Quantum supremacy using a programmable superconducting processor},
journal={Nature},
year={2019},
month={Oct},
day={01},
volume={574},
number={7779},
pages={505-510},
issn={1476-4687},
doi={10.1038/s41586-019-1666-5},
url={https://doi.org/10.1038/s41586-019-1666-5}
}

@article{Liu23,
  title = {Single Flux Quantum-Based Digital Control of Superconducting Qubits in a Multichip Module},
  author = {Liu, C.H. and Ballard, A. and Olaya, D. and Schmidt, D.R. and Biesecker, J. and Lucas, T. and Ullom, J. and Patel, S. and Rafferty, O. and Opremcak, A. and Dodge, K. and Iaia, V. and McBroom, T. and DuBois, J.L. and Hopkins, P.F. and Benz, S.P. and Plourde, B.L.T. and McDermott, R.},
  journal = {PRX Quantum},
  volume = {4},
  issue = {3},
  pages = {030310},
  numpages = {15},
  year = {2023},
  month = {Jul},
  publisher = {American Physical Society},
  doi = {10.1103/PRXQuantum.4.030310},
  url = {https://link.aps.org/doi/10.1103/PRXQuantum.4.030310}
}

@article{Jiang25,
  title = {On-chip direct-current source for scalable superconducting quantum computing},
  author = {Jiang, Lei and Xu, Yu and Li, Shaowei and Yan, Zhiguang and Gong, Ming and Rong, Tao and Sun, Chenyin and Sun, Tianzuo and Jiang, Tao and Deng, Hui and Zha, Chen and Lin, Jin and Chen, Fusheng and Zhu, Qingling and Ye, Yangsen and Rong, Hao and Yan, Kai and Cao, Sirui and Li, Yuan and Guo, Shaojun and Qian, Haoran and Hu, Yisen and Wu, Yulin and Li, Yu-Huai and Wu, Gang and Wang, Xueshen and Wang, Shijian and Cao, Wenhui and Wang, Yeru and Huo, Yong-Heng and Li, Jinjin and Peng, Cheng-Zhi and Zhu, Xiaobo and Pan, Jian-Wei},
  journal = {Phys. Rev. Appl.},
  volume = {24},
  issue = {3},
  pages = {034057},
  numpages = {9},
  year = {2025},
  month = {Sep},
  publisher = {American Physical Society},
  doi = {10.1103/qwvv-mz8s},
  url = {https://link.aps.org/doi/10.1103/qwvv-mz8s}
}

@article{Pechal16,
  title = {Superconducting Switch for Fast On-Chip Routing of Quantum Microwave Fields},
  author = {Pechal, M. and Besse, J.-C. and Mondal, M. and Oppliger, M. and Gasparinetti, S. and Wallraff, A.},
  journal = {Phys. Rev. Appl.},
  volume = {6},
  issue = {2},
  pages = {024009},
  numpages = {8},
  year = {2016},
  month = {Aug},
  publisher = {American Physical Society},
  doi = {10.1103/PhysRevApplied.6.024009},
  url = {https://link.aps.org/doi/10.1103/PhysRevApplied.6.024009}
}

@misc{Reitz26,
      title={Transmon Phase Gates Controlled by Superconducting Soliton DAC}, 
      author={Derek Reitz and Tony X. Zhou and Aditya Sharma and Ryan Bilotta and John McFarland and Aref Fouladi and Jacob Glasby and Aruna Ramanayaka and Zachary Stegen and Aaron Pesetski and Mark Covington and Gregory Boyd and Jeremy Clark},
      year={2026},
      eprint={2607.05072},
      archivePrefix={arXiv},
      primaryClass={quant-ph},
      url={https://arxiv.org/abs/2607.05072}, 
}


\end{document}